\documentclass[12pt]{article}
\usepackage{amsmath,amssymb}
\usepackage{latexsym,graphicx,color}

\def\6#1{{\underline{#1}}}
\def\m6#1{{\underline{#1}\,}}

\newdimen\Tdim
\def\ispan{{\setbox0=\hbox{i}%
\Tdim\ht0\advance\Tdim\dp0\rule[-\dp0]{0pt}{\Tdim}}}
\def\jspan{{\setbox0=\hbox{j}%
\Tdim\ht0\advance\Tdim\dp0\rule[-\dp0]{0pt}{\Tdim}}}
\def\Tspan#1{{\setbox0=\hbox{#1}%
\Tdim\ht0\advance\Tdim\dp0\advance\Tdim.55ex\rule[-\dp0]{0pt}{\Tdim}\box0}}

\def\be{\begin{eqnarray}}
\def\ben{\begin{eqnarray*}}
\def\ee{\end{eqnarray}}
\def\een{\end{eqnarray*}}
\def\Tr{{\rm Tr}}

\def\p{\partial}

\def\D{\mathcal{D}}

\def\=:{=\hspace{-.7em}\raisebox{1.1ex}{.}\hspace{.1em}\raisebox{-0.2ex}{.} }
\newcommand{\ibf}[1]{\mbox {\boldmath{$ #1 $}}}
\newcommand{\NF}{N_{\rm F}}
\newcommand{\NC}{N_{\rm C}}
\newcommand{\hs}[1]{\hspace{#1 mm}}

\newcommand {\beq}{\begin{eqnarray}}
\newcommand {\eeq}{\end{eqnarray}}

\makeatletter
\@addtoreset{equation}{section}

\renewcommand{\thefootnote}{\fnsymbol{footnote}}
\makeatother

\makeatletter
\newcommand{\thetablename}{Table}
\def\fnum@table{\thetablename\ \thetable}
\makeatother

\begin{document}
\thispagestyle{empty}
\begin{flushright}
IFUP-TH/17, TIT/HEP-573, \\
{\tt hep-th/} \\
Jul, 2007 \\
\end{flushright}
\vspace{3mm}
\begin{center}
{\LARGE \bf 
Dynamics of Domain Wall Networks
} \\ 
\vspace{5mm}

{\normalsize\bfseries
Minoru~Eto$^{a,b}$, 
Toshiaki~Fujimori$^c$, 
Takayuki Nagashima$^c$, \\
Muneto~Nitta$^d$,
Keisuke~Ohashi$^e$, 
and
 Norisuke~Sakai$^c$}
\footnotetext{
e-mail~addresses: \tt
minoru@df.unipi.it;
fujimori,
nagashi@th.phys.titech.ac.jp;\\
nitta@phys-h.keio.ac.jp;
K.Ohashi@damtp.cam.ac.uk;
nsakai@th.phys.titech.ac.jp
}

\vskip 1.5em
$^a$ {\it INFN, Sezione di Pisa,
Largo Pontecorvo, 3, Ed. C, 56127 Pisa, Italy
}
\\
$^b$ {\it Department of Physics, University of Pisa
Largo Pontecorvo, 3,   Ed. C,  56127 Pisa, Italy
}
\\
$^c$ {\it Department of Physics, Tokyo Institute of
Technology, Tokyo 152-8551, Japan
}
\\
$^d$ 
{\it Department of Physics, Keio University, Hiyoshi, Yokohama,
Kanagawa 223-8521, Japan
}
\\
$^e$
{\it Department of Applied Mathematics and Theoretical Physics, \\
University of Cambridge, CB3 0WA, UK}
\vspace{12mm}

\abstract{
Networks or webs of domain walls are admitted in 
Abelian or non-Abelian gauge theory coupled to 
fundamental Higgs fields with complex masses. 
We examine the dynamics of the domain wall loops 
by using the moduli approximation and find 
a phase rotation induces a repulsive force 
which can be understood as a Noether charge of $Q$-solitons. 
Non-Abelian gauge theory 
allows different types of loops 
which can be deformed to each other 
by changing a modulus. 
This admits the moduli geometry like a sandglass 
made by gluing the tips of the two cigar-(cone-)like metrics 
of a single triangle loop. 
We conclude that the sizes of all loops tend 
to grow for a late time in general models 
with complex Higgs masses, 
while the sizes are stabilized at some values 
once triplet masses are introduced for the Higgs fields.
We also show that the stationary motion on the moduli space 
of the domain wall webs represents 1/4 BPS $Q$-webs of walls.

}

\end{center}

\vfill
\newpage
\setcounter{page}{1}
\setcounter{footnote}{0}
\renewcommand{\thefootnote}{\arabic{footnote}}

\section{Introduction}\label{sec:intro}

In various area of physics, 
many kinds of topological defects are expected to be 
produced at a phase transition 
via the Kibble mechanism \cite{Kibble:1976sj}. 
More than two extended objects like cosmic strings or domain walls 
intersect or meet with angles in general, 
and therefore such a production inevitably results in 
networks or webs of these objects \cite{VS}. 
In the condensed matter physics several examples have been observed 
while it is not the case of particle physics, 
astrophysics or cosmology. 
Previously cosmic string junctions were suggested 
to be a seed of galaxy formation. 
Although such a possibility has been denied by 
a recent cosmic microwave background data, 
it is argued that they may still play a certain role. 
A domain wall network was proposed to explain 
dark matter/energy \cite{dark-energy}. 
Future observation of such defect networks 
in our Universe certainly deserves to be explored. 
Usually dynamics of these networks have been 
studied by computer simulation. 
On the other hand, 
in the case of particle-like solitons 
such as monopoles, the analytic method of 
the moduli space (geodesic) 
approximation has been developed \cite{Manton:1981mp,Manton:2004tk}.
By this their low energy dynamics can be described as 
geodesics of the moduli space of these solitons. 
Therefore the determination of the moduli space 
is crucial for this task. 
In a previous paper \cite{Eto:2006bb} 
we have successfully constructed 
the moduli space of domain wall networks in a certain 
model which allows a supersymmetric generalization. 
Supersymmetry is expected to exist in the early Universe 
so this situation is realistic. 

In this paper we will work out the dynamics of 
domain wall networks using the moduli space approximation. 
We find that the sizes of all loops tend 
to grow for a late time in general models 
with complex Higgs masses, 
while the sizes are stabilized at some values 
once triplet masses are introduced for the Higgs fields.
To the best of our knowledge 
this is the first example to discuss 
the dynamics of a composite system of solitons 
analytically.\footnote{
We have analyzed analytically 
compressed walls which appear as limiting 
configurations of multiple parallel walls compressed to each 
other\cite{Isozumi:2004jc}. 
Although these configurations may be regarded as composite 
solitons, they can be obtained as a smooth limiting 
point within a moduli space of multiple parallel walls, and 
do not present qualitatively new features unlike our 
present case of the 1/4 BPS webs of walls. 
} 
Our model here 
is a $U(N_{\rm C})$ gauge theory 
coupled to $N_{\rm F}$ Higgs fields 
in the fundamental representation, 
which can be extended to possess 
${\cal N}=2$ supersymmetry. 
This model has been recently studied extensively 
because it allows many kinds of 
Bogomol'nyi-Prasad-Sommerfield (BPS) 
solitons, 
see \cite{Tong:2005un,Eto:2006pg,Shifman:2007ce} for a review.
Vacua are isolated and disconnected in theories with 
the number of flavors more than the number of color, 
$N_{\rm F} > N_{\rm C}$, and 
with non-degenerate masses for the Higgs fields \cite{Arai:2003tc}.
Parallel multiple domain wall solutions exist 
as 1/2 BPS states 
when the Higgs masses are real and non-degenerate. 
By introducing the method of the moduli matrix 
\cite{Eto:2006pg,rev-modulimatrix},
{\it analytic} solutions of these domain walls were constructed 
in strong gauge coupling limit 
\cite{Isozumi:2004jc}
(see \cite{U(1)walls} for domain walls in 
$U(1)$ gauge theory). 
This method was then applied to construct 
vortex solutions \cite{Eto:2005yh}, 
vortex-strings stretched between 
parallel domain walls \cite{Isozumi:2004vg}, 
instantons inside a vortex-sheet \cite{Eto:2004rz}.    
Finally the most general {\it analytic} solutions of 
1/4 BPS networks (webs) of 
domain walls have been constructed 
in models with complex non-degenerate Higgs masses 
\cite{Eto:2005cp,Eto:2005fm,Eto:2005mx}.\footnote{
In ${\cal N}=1$ supersymmetric field theories, 
junctions of domain walls 
were previously found to be 1/4 BPS
states preserving only a quarter of supersymmetry  
\cite{Gibbons:1999np}. 
Exact solutions of wall junctions were constructed 
in \cite{Oda:1999az}.
See \cite{Eto:2006bb} for more complete references 
of domain wall junctions in 
${\cal N}=1$ supersymmetric field theories.
}
These solutions contain full moduli of 
a network with arbitrary numbers 
of loops and external legs of walls. 
Effective K\"ahler potential of 1/2 BPS solitons 
was constructed 
in the superfield formalism \cite{Eto:2006uw} 
and then it has been generalized to the case of 
domain wall networks \cite{Eto:2006bb}.
The zero modes of external legs are 
non-normalizable and have to be fixed to discuss dynamics, 
while zero modes corresponding to loop size and 
associated internal phase are normalizable 
and appear as massless fields in the effective theory. 
We have constructed the effective K\"ahler potential 
and the metric of the simplest triangle loop 
in $U(1)$ gauge theory coupled 
with $N_{\rm F}=4$ Higgs scalars 
and have found that the metric has a geometry 
between a cone and a cigar \cite{Eto:2006bb}. 
This metric is rather non-trivial since 
it is regular on the tip although it corresponds to shrinking loop. 
Therefore it is expected to describe a smooth bounce of the loop.

In this paper we discuss the dynamics 
of loops of domain walls for 
1) a triangle single loop in the simplest model of 
$N_{\rm C}=1$ and $N_{\rm F} = 4$, 
2) a double loop in the model with 
$N_{\rm C}=1$ and $N_{\rm F} = 6$ 
and 3) a non-Abelian loop 
in the model with $N_{\rm C}=2$ and $N_{\rm F} = 4$.  
This paper is organized as follows. 
In section 2 we summarize the previous results 
on the construction of domain wall networks and 
the effective action on them. 
In section 3
we first investigate the dynamics of the single triangle loop.  
The moduli metric allows the $U(1)$ isometry 
whose orbit is parametrized by 
a Nambu-Goldstone mode of the flavor symmetry 
spontaneously broken by the configuration. 
Associated with this isometry, 
a conserved charge $Q$ exists in
the general motion of the moduli space.  
When $Q=0$, a motion of a shrinking loop is bounced 
and the phase is rotated with the angle $\pi$
after the loop completely shrinks. 
When $Q\neq 0$ a shrinking loop is bounced 
at the minimum size of the loop determined by $Q$.
In section 4 we investigate the dynamics of double loop.
In this case there exist two conserved charges 
$Q_1$ and $Q_2$ correspond to the phases of the two loops. 
Both loops will grow after their sizes bounce at the minimum
irrespective of their $Q$-charges.
In section 5 we work out the dynamics of non-Abelian loop. 
In this case there exist two different configurations of 
non-planar webs with a non-Abelian loop, 
which can be deformed to each other by changing a modulus. 
The moduli space geometry looks like a sandglass 
which is made by gluing the tips of the 
two metrics of a single triangle loop. 
Each region of the sandglass metric corresponds to 
the configuration of each non-Abelian loop. 
Depending on the value of the conserved charge $Q$ 
one configuration can or cannot change to 
the other configuration. 
In section 6 
we turn on the triplet masses for the Higgs fields. 
In the context of field theory with eight supersymmetry charges 
this is possible in three space-time dimensions. 
We find the third masses induce the attractive force 
for the loop size whereas
the $Q$-charge induces the repulsive force. 
Then the size of the loop is stabilized at some value 
where two kinds of forces are balanced.
This mechanism to stabilize the size is the same 
as the one of the $Q$-lumps in nonlinear sigma models 
with a potential term \cite{Leese:1991hr}--\cite{Eto:2005sw}; 
the size of $Q$-lumps are stabilized 
by the $Q$-charge and the masses.
Also, it was shown in \cite{Tong:1999mg} that 
1/4 BPS dyon can be understood as 
stationary motion in the moduli space 
of BPS monopoles with a potential term 
induced by the masses. 
Dyonic instanton is also 
understood as stationary motion 
in the moduli space of instantons \cite{Lambert:1999ua}.
In the same way, our motion in the moduli space 
of the domain wall webs 
suggests BPS dyonic extension of domain wall webs. 
In fact 
it has been previously shown in \cite{Eto:2005sw} 
that the configuration of $Q$-domain wall webs 
is again 1/4 BPS (but not 1/8 BPS) and is stable.
We reexamine this interpretation in this section.  
Section 7 is devoted to Conclusion and Discussion.
Implication of our work to cosmology is briefly discussed.

\section{Effective Action of Domain Wall Networks}

\subsection{BPS Equations for Domain Wall Networks}

Let us here briefly present our model 
(see \cite{Eto:2006pg} for a review), which admits 
1/4 BPS webs of domain walls. 
We consider 3+1 dimensional 
$\mathcal{N}=2$ supersymmetric $U(\NC)$ gauge theory 
with $\NF\,\,(>\NC)$ massive hypermultiplets 
in the fundamental representation. 
Here the bosonic components in the vector multiplet are 
gauge fields $W_M~(M=0,1,2,3)$, 
the real scalar fields $\Sigma_\alpha~(\alpha=1,2)$ 
in the adjoint representation, 
and those in the hypermultiplet are 
the $SU(2)_R$ doublets of the complex scalar fields $H^i~(i=1,2)$, 
which we express as $\NC \times \NF$ matrices. 
After eliminating the auxiliary fields, we obtain the 
bosonic part of the Lagrangian as 
\beq
\mathcal{L} &=& 
\Tr \left[-\frac{1}{2g^2}F_{MN}F^{MN} 
+ \frac{1}{g^2} \sum_{\alpha=1}^2 
\mathcal D_M \Sigma_\alpha \mathcal D^M \Sigma_\alpha
+ \mathcal D_M H^i (\mathcal D^M H^i)^\dag \right] 
- V, \label{eq:lag}\\
V &=& 
\Tr \left[ \frac{1}{g^2} \sum_{a=1}^3 (Y^a)^2 
+ \sum_{\alpha=1}^2 
(H^iM_\alpha-\Sigma_\alpha H^i)(H^iM_\alpha-\Sigma_\alpha H^i)^\dag
- \frac{1}{g^2}[\Sigma_1,\Sigma_2]^2 \right],
\label{eq:pot}
\eeq
where we have defined 
$Y^a \equiv \frac{g^2}{2} \left(c^a \mathbf 1_{\NC} - {(\sigma^a)^j}_i H^i(H^j)^\dagger \right)$ 
with $g$ the gauge coupling for $U(N_{\rm C})$ gauge theory, and 
$c^a$ an $SU(2)_R$ triplet of the Fayet-Iliopoulos (FI) parameters. 
In the following, we choose the FI parameters as $c^a=(0,0,c>0)$ 
by using $SU(2)_R$ rotation without loss of generality. 
Here we use the space-time metric $\eta_{MN}=\text{diag}(+1,-1,-1,-1)$ 
and $M_\alpha$ are real diagonal mass matrices, 
$M_1=\text{diag} (m_1,m_2,\cdots,m_{\NF})$, 
$M_2=\text{diag}(n_1,n_2,\cdots,n_{\NF})$. 
The covariant derivatives are defined as 
$\mathcal D_M \Sigma =\partial_M \Sigma + i[W_M,\Sigma],$ 
$\mathcal D_M H^i=(\partial_M+iW_M)H^i$, 
and the field strength is defined as 
$F_{MN}=-i[\mathcal D_M,\mathcal D_N]=\partial_MW_N-\partial_NW_M +i[W_M,W_N]$.

If we turn off all the mass parameters, 
the vacuum manifold is 
the cotangent bundle over the complex Grassmannian 
$T^\ast Gr_{\NF,\NC}$ \cite{Lindstrom:1983rt}.  
Once the mass parameters $m_A + i n_A,~(A=1,\cdots \NF)$ 
are turned on and chosen to be fully non-degenerate 
($m_A+in_A \neq m_B + in_B$ for $A\neq B$),
the almost all points of the vacuum manifold 
are lifted and only 
${}_{\NF} C_{\NC}=\NF!/\left[\NC!(\NF-\NC)!\right]$ 
discrete points on the base manifold $Gr_{\NF,\NC}$ 
are left to be the supersymmetric vacua \cite{Arai:2003tc}. 
This choice of the mass parameters breaks 
the $SU(\NF)$ flavor symmetry to $U(1)^{\NF-1}$. 
Each vacuum is characterized 
by a set of $\NC$ different indices 
$\left< A_1,\cdots,A_{\NC}\right>,\,1\leq A_1 < \cdots < A_{\NC} \leq \NF$, 
which we will often abbreviate 
as $\langle A_r \rangle$ in the following.
In these vacua, the vacuum expectation values are determined as
\beq
\left<H^{1rA}\right> = \sqrt{c} \, {\delta^{A_r}}_A, \hs{5} 
\left<H^{2rA}\right> = 0, \hs{5} 
\left<\Sigma\right> = {\rm diag} \left(m_{A_1}+in_{A_1},\cdots,m_{A_{\NC}}+in_{A_{\NC}}\right), 
\label{vacuum}
\eeq
where $r$ is color index running from 1 to $\NC$, 
the flavor index $A$ runs from 1 to $\NF$ and
$\Sigma$ is the complex adjoint scalar 
defined by $\Sigma \equiv \Sigma_1 + i \Sigma_2$. 

The 1/4 BPS equations for webs of walls 
interpolating the discrete vacua (\ref{vacuum}) 
can be obtained by usual Bogomol'nyi completion  
of the energy density as \cite{Eto:2005cp,Eto:2005fm} 
\begin{align}
&F_{12}=i[\Sigma_1,\Sigma_2],~~~~\mathcal D_1\Sigma_2=\mathcal D_2\Sigma_1,~~~~
\mathcal D_1\Sigma_1+\mathcal D_2\Sigma_2=Y^3,\label{eq:BPS2} \\
&\mathcal D_1H^1=H^1M_1-\Sigma_1H^1,~~~~\mathcal D_2H^1=H^1M_2-\Sigma_2H^1. \label{eq:BPS1}
\end{align}
Here we consider static configurations 
which are independent of $x_3$, 
so we set $\partial_0=\partial_3=0$ and $W_0=W_3=0$. 
Furthermore, we take $H^2=0$ 
because it always vanishes for the 1/4 BPS solutions. 
The Bogomol'nyi energy bound is given by
\begin{align}
\mathcal{E} 
\geq \mathcal{Y}+\mathcal{Z}_1+\mathcal{Z}_2+\partial_\alpha J_\alpha,
\label{eq:Bogomolny}
\end{align}
where the central (topological) charge densities 
which characterize the solutions are of the form
\begin{align}
\mathcal{Y}=\frac{2}{g^2}\partial_\alpha \text{Tr}(\epsilon^{\alpha \beta}\Sigma_2 \mathcal D_\beta \Sigma_1),~~~~~~
\mathcal{Z}_1=c\,\partial_1 \text{Tr} \Sigma_1,~~~~~~
\mathcal{Z}_2=c\,\partial_2 \text{Tr} \Sigma_2.
\end{align}
The topological charges are defined by
\beq
T_{\rm w} \equiv \int d^2 x \left( \mathcal Z_1 + \mathcal Z_2 \right),\hs{10}
Y \equiv \int d^2 x \, \mathcal Y.
\label{topo}
\eeq
Here $T_{\rm w}$ corresponds to the energy of domain walls 
and $Y$ corresponds to the energy of domain wall junctions. 
Since energy of domain walls means tension times the length of the walls, 
this quantity is divergent. 
On the other hand $Y$ has a finite value, 
and we call this charge as the junction charge or the Hitchin charge.
Note that the integration of the fourth term
$\partial_\alpha J_\alpha=\partial_\alpha \text{Tr}[H^1(M_\alpha H^\dag-H^{1\dag} \Sigma_\alpha)]$
in Eq.\,(\ref{eq:Bogomolny}) 
does not contribute to the topological charges. 

The 1/4 BPS equations Eq.\,(\ref{eq:BPS2}) and 
Eq.\,(\ref{eq:BPS1})~\cite{Eto:2005cp,Eto:2005fm} 
can be solved as follows.
Firstly, since the first two equations in Eq.\,(\ref{eq:BPS2}) 
give an integrability condition for the two operators
${\cal D}_\alpha + \Sigma_\alpha\ (\alpha=1,2)$, 
$W_\alpha$ and $\Sigma_\alpha$ can be written as 
\begin{align}
\Sigma_\alpha+iW_\alpha=S^{-1}\partial_\alpha S.
\label{eq:solBPS1}
\end{align}
Here, $S(x^1,x^2) \in GL(\NC,{\bf C})$ is a matrix valued function.
Secondly, Eq.\,(\ref{eq:BPS1}) can be solved as 
\begin{align}
H^1=S^{-1}H_0 \, e^{M_1x^1+M_2x^2}.
\label{eq:solBPS2}
\end{align}
Here $H_0$, which we call ``moduli matrix", 
is an $\NC \times \NF$ constant complex matrix of rank $\NC$, 
and contains all the moduli parameters of solutions. 
Any sets of $S$ and moduli matrix $H_0$ 
related by the following $V$-transformation 
are physically equivalent 
since they do not change the physical configuration:
\begin{align}
H_0 \rightarrow VH_0,~~~~S(x^1,x^2) \rightarrow VS(x^1,x^2),~~~~V \in GL(\NC, {\bf C}).
\end{align}
Finally, the last equation in Eq.\,(\ref{eq:BPS2}) can be converted, 
by using an $\NC \times \NC$ matrix valued function defined by 
\beq
\Omega(x^1,x^2) \equiv SS^\dagger, 
\eeq
to the following equation:
\beq
\frac{1}{cg^2}\bigl(\partial_\alpha(\Omega^{-1}\partial_\alpha \Omega)\bigl)=
\ibf{1}_{\NC}-\Omega^{-1}\Omega_0,
\label{eq:master}
\eeq
where $\Omega_0 \equiv \frac{1}{c} H_0 \, e^{2(M_1x^1+M_2x^2)} H_0^\dag$. 
This equation is called the master equation for webs of walls. 
Since $HH^\dagger - c \mathbf 1_{\NC}=0$ in vacuum regions, the solution $\Omega(x^1,x^2)$ of the master equation should approach $\Omega_0$ near the vacuum regions. 
It determines $S$ for a given moduli matrix $H_0$ 
up to gauge transformations 
and then the physical fields can be obtained 
through Eq.\,(\ref{eq:solBPS1}) and Eq.\,(\ref{eq:solBPS2}). 

\begin{figure}[h]
\begin{center}
\begin{tabular}{ccc}
\includegraphics[height=4cm]{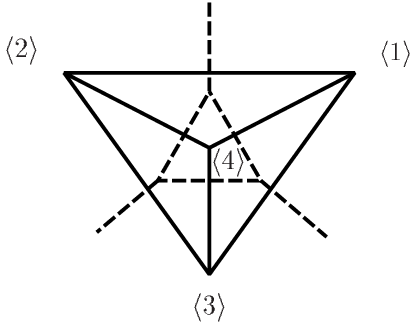}&\qquad&
\includegraphics[height=4cm]{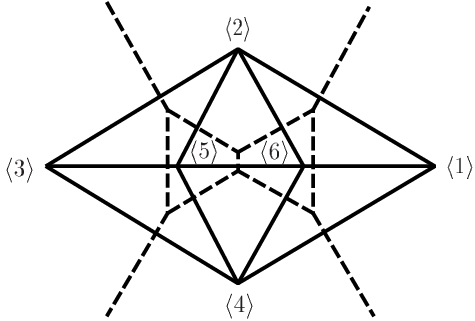}\\
{\small (a) triangle loop in $\NC=1$, $\NF=4$ model}&\qquad&
{\small (b) double loop in $\NC=1$, $\NF=6$ model} 
\end{tabular}
\caption{\small Grid diagram and web diagram in Abelian gauge theory.}
\label{fig:ab-grid}
\end{center}
\end{figure}

\begin{figure}[htb]
\begin{center}
\begin{tabular}{ccc}
\includegraphics[height=2.9cm]{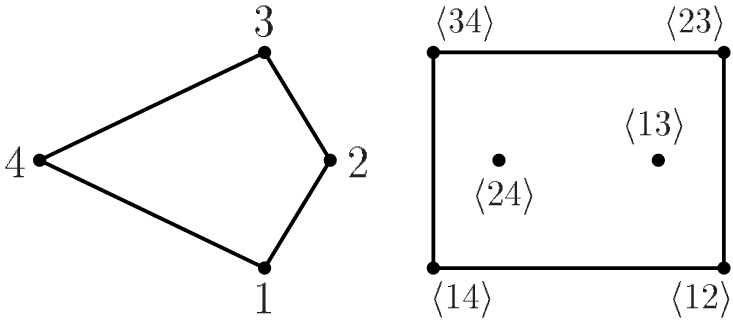}&\quad&
\includegraphics[height=3.5cm]{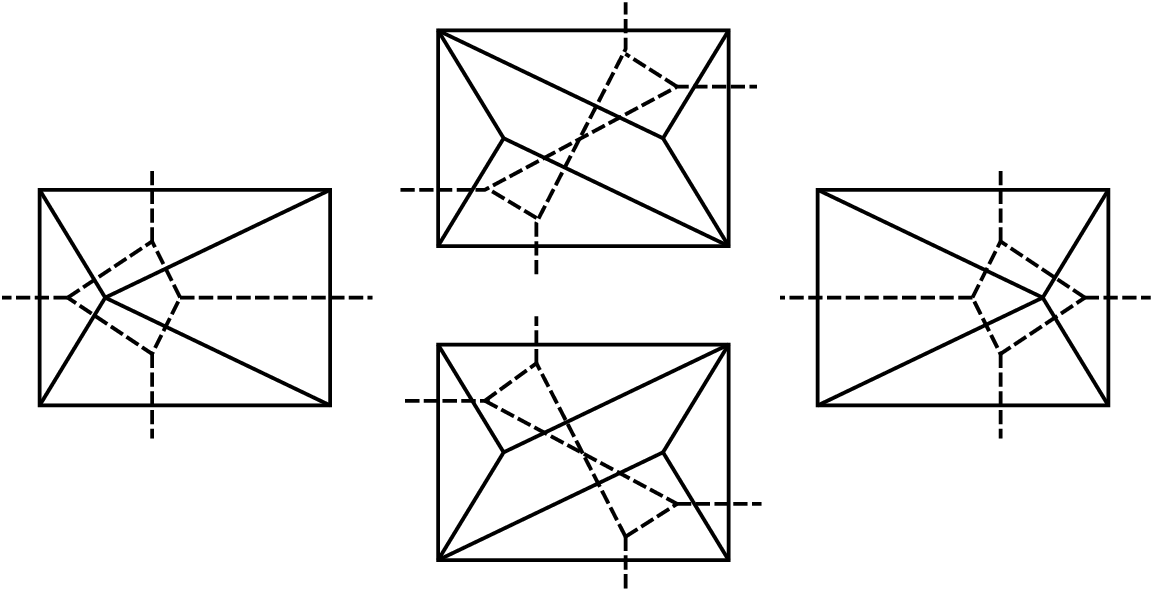}\\
{\small (a) mass arrangement and vacuum points} &\quad&
{\small (b) non-Abelian loop in $\NC=2$, $\NF=4$ model }
\end{tabular}
\caption{\small Grid diagram and web diagram in non-Abelian gauge theory.}
\label{fig:nab-grid}
\end{center}
\end{figure}

There is a useful diagram to understand the structure of webs of walls, 
which is called the grid diagram \cite{Eto:2005cp,Eto:2005fm}.
The grid diagram is a convex polygon 
in the complex plane $\Tr \left< \Sigma \right> (\Sigma \equiv \Sigma_1+i\Sigma_2)$.
A vacuum point labeled by $\langle A_1 \cdots A_{\NC} \rangle$ 
correspond to the vertex of the convex polygon plotted at 
$\Tr\left<\Sigma\right> = \sum_{r=1}^{\NC} \left( m_{A_r} + in_{A_r} \right)$.
For each edge connecting two vertices, 
there is a domain wall interpolating the two vacua 
and each triangle corresponds to a 3-pronged domain wall junction.
Some examples of grid diagrams are shown 
in Fig.\,\ref{fig:ab-grid} and Fig.\,\ref{fig:nab-grid}. 
In non-Abelian gauge theory, 
two vacua with only one different label 
such as $\left<\ \dots \ A\right>$ 
and $\left<\ \dots \ B\right>$ can be connected 
while two with $\left<\ \dots \ AB\right>$ 
and $\left<\ \dots \ CD\right>$ are forbidden to be connected. 
If there are several ways to connect the vacuum points, 
we obtain different configurations 
as shown in Fig.\,\ref{fig:nab-grid}.
By varying the moduli parameters, 
we can move one configuration to another one. 

One can easily read physical informations 
about domain walls and junctions from the grid diagram.
The tension of the domain wall is proportional 
to the length of the corresponding edge of the grid diagram. 
More precisely, for a domain wall interpolating 
between vacuum $\langle ...  A \rangle$ 
and vacuum $\langle ...  B \rangle$, 
the tension is given by
\beq
T^{\langle ...  A \rangle \langle ...  B \rangle}=c\,|\vec m_A-\vec m_B|,
\label{eq:tension}
\eeq
where $\vec m_A,\,\vec m_B$ are two component vectors 
such that $\vec m_A = \left(m_A,n_A \right)$, 
$\vec m_B = \left(m_B,n_B\right)$. 
Furthermore the magnitude of the junction charge is 
proportional to the area of the corresponding triangle 
and its sign can be read off from the vacuum labels.
If the junction interpolates three different vacua 
with labels such as $\langle ...  A \rangle 
\langle ...  B \rangle \langle ...  C \rangle$,
this junction is called ``Abelian junction" 
and its junction charge is given by
\beq
Y^{\langle ...  A \rangle \langle ...  B \rangle
\langle ...  C \rangle}=-\frac{|\Delta_{[ABC]}|}{g^2},
\label{eq:abelian}
\eeq
where we have defined $\Delta_{[ABC]}$ as
\beq
\Delta_{[ABC]}
=\vec m_A\times \vec m_B +\vec m_B \times \vec m_C 
+\vec m_C \times \vec m_A,
\label{eq:delta}
\eeq
which is twice the area of the triangle in the grid diagram. 
The junction charge above is negative, 
and can be interpreted as the binding energy 
of domain walls at the junction point. 
On the other hand, if the junction interpolates three vacua 
with labels such as $\langle ...  AB \rangle 
\langle ...  BC \rangle \langle ...  AC \rangle$, 
this junction is called ``non-Abelian junction" 
and its junction charge is given by
\beq
Y^{\langle ...  AB \rangle \langle ...  BC \rangle
\langle ...  CA \rangle}=\frac{|\Delta_{[ABC]}|}{g^2}.
\label{eq:non-abelian}
\eeq
This is positive, 
and can be interpreted as the 
Hitchin charge of the Hitchin system. 
The details can be seen in \cite{Eto:2005fm}. 

In order to extract concrete informations from the moduli matrix $H_0$, 
it is useful to denote 
$\det H_0^{\left<A_r\right>} = \exp{\left(a^{\left<A_r\right>} + ib^{\left<A_r\right>}\right)}$, 
where $H_0^{\left<A_r\right>}$ is an 
$\NC \times \NC$ minor matrix 
whose elements are given by 
$(H_0^{\left<A_r\right>})^{st} = \left(H_0\right)^{sA_t}$.
Defining the weight ${\cal W}^{\left<A_r\right>}$ of the vacuum 
$\left<A_r\right> = \left<A_1A_2\cdots A_{\NC}\right>$ by 
\beq
{\cal W}^{\left<A_r\right>} (x^1,x^2) \equiv
\sum_{r=1}^{\NC} \left(m_{A_r}x^1 + n_{A_r}x^2\right) 
+ a^{\left<A_r\right>},
\label{eq:vacuum_weight}
\eeq
we can write the determinant of $\Omega_0$ as 
\begin{equation}
\det \Omega_0 
=\det\left(\frac{1}{c} H_0 \, e^{2(M_1x^1+M_2x^2)} H_0^\dag \right)
=\frac{1}{c^{\NC}}\sum_{\langle A_r \rangle}e^{2{\cal W}^{\langle A_r \rangle}}.\label{eq:omega0_vacuum_weight}
\end{equation}
If only one of the weight $\mathcal W^{\langle A_r \rangle}$ 
is non-zero, 
we can show that the configuration is 
the vacuum labeled by $\langle A_r \rangle$. 
Since the solution of the master equation $\Omega$ is 
well-approximated by $\Omega_0$ near vacuum regions, 
we can estimate the position of the domain wall
interpolating between vacuum $\langle A_r \rangle$ 
and vacuum $\langle B_r \rangle$ 
as a line on which the weights 
$\mathcal W^{\langle A_r \rangle}$ and 
$\mathcal W^{\langle B_r \rangle}$ are comparable:
\beq
\mathcal W^{\langle A_r \rangle} - 
\mathcal W^{\langle B_r \rangle} = 
\sum_{r=1}^{\NC}\left(m_{A_r}-m_{B_r}\right) x^1 +
\sum_{r=1}^{\NC}\left(n_{A_r}-n_{B_r}\right) x^2 +
a^{\left<A_r\right>} - a^{\left<B_r\right>} \simeq 0.
\label{eq:posi-ang}
\eeq
Here the other weights should be sufficiently smaller than 
$\mathcal W^{\langle A_r \rangle}$ and 
$\mathcal W^{\langle B_r \rangle}$ .
Hence the parameter 
$a^{\left<A_r\right>} - a^{\left<B_r\right>}$ 
in the moduli matrix determines 
the position of the domain wall. 
Furthermore, one can see the angle of the domain wall
is determined by the mass difference between the two vacua.
Notice that the domain wall line Eq.\,(\ref{eq:posi-ang}) is 
perpendicular to the corresponding 
edge of the grid diagram, see Fig.\,\ref{fig:ab-grid}. 
So the grid diagram gives us informations of 
the shape of the domain wall web as a dual diagram. 
A junction point at which three of domain walls get together 
can also be estimated by the condition of 
equating the weights of three related vacua as 
${\cal W}^{\left<A_r\right>} 
\simeq {\cal W}^{\left<B_r\right>} 
\simeq {\cal W}^{\left<C_r\right>}$.

\subsection{Effective Action of Domain Wall Networks}

Once we obtain the solutions of the BPS equations 
Eq.\,(\ref{eq:BPS2}) and Eq.\,(\ref{eq:BPS1}), 
we can construct 
a low-energy effective theory on the 
world-volume of the domain wall networks. 
While all the massive modes on the background BPS solutions 
can be ignored at low-energies, 
moduli parameters (zero modes) as elements of the moduli 
matrix $H_0$ can provide massless modes which will 
play a main role in the effective theory. 
Among these zero modes, we should promote only 
normalizable zero modes $\phi_i$ 
to fields on the world-volume of the domain wall network as 
\beq
H_0\bigl(\phi^i \bigr) \rightarrow H_0 \bigl(\phi^i(x^\mu)\bigr),
\label{eq:prom}
\eeq
where $x^\mu \, (\mu=0,3)$ denotes the world-volume coordinates 
of the domain wall network. 
On the other hand, non-normalizable zero modes 
which change the boundary conditions 
at spatial infinities cannot be promoted 
to fields on the world-volume. 

In general, the master equation Eq.\,(\ref{eq:master}) is difficult to solve. 
However we can obtain a general form of the effective Lagrangian 
for the moduli fields without solving the master equation, 
which have been constructed in \cite{Eto:2006bb}.
It was found that the metric on the moduli space 
is a K\"ahler metric whose K\"ahler potential is given by 
\begin{eqnarray}
K(\phi,\bar \phi)
= \int d^2x\, \left[c\,{\rm log~det}\,\Omega_{\rm sol}(\phi,\bar\phi)
+\frac{1}{2g^2}{\rm Tr}(\Omega^{-1}_{\rm sol}(\phi,\bar\phi)\partial_\alpha 
\Omega_{\rm sol}(\phi,\bar\phi))^2\right],
\label{eq:kah}
\end{eqnarray}
where $\Omega_{\rm sol}(\phi,\bar\phi)$ is a solution of 
the master equation (\ref{eq:master}). 
In order to get this K\"ahler potential, one needs to solve 
the Gauss's law constraint 
for the world-volume elements of the gauge field $W_\mu(x^\mu)$
\beq
\D_\alpha F_{\alpha 0} - i\left[\Sigma_\alpha,\D_0\Sigma_\alpha\right]
- i \frac{g^2}{2}\left(H^1\D_0H^{1\dagger} 
- \D_0 H^1 H^{1\dagger}\right) = 0.
\eeq
We found \cite{Eto:2006bb} a generic form of the solution 
for the Gauss's law can be 
expressed by derivatives  with respective to the moduli 
fields as
\beq
W_\mu(x^\mu) = i \left( \delta_\mu 
S_{\rm sol}^\dagger(\phi,\bar\phi) 
S_{\rm sol}^{\dagger -1}(\phi,\bar\phi) 
- S_{\rm sol}^{-1}(\phi,\bar\phi) 
\delta_\mu^\dagger S_{\rm sol}(\phi,\bar\phi)\right),
\label{eq:sol_Gauss}
\eeq
where $S_{\rm sol}(\phi,\bar\phi)$ is given by 
$\Omega_{\rm sol}(\phi,\bar\phi) 
= S_{\rm sol}(\phi,\bar\phi)S_{\rm sol}(\phi,\bar\phi)^\dagger$
and the variations are defined by 
$\delta_\mu = \partial_\mu \phi^i \frac{\partial}{\partial \phi^i}$ and
$\delta_\mu^\dagger = \partial_\mu \bar \phi^i \frac{\partial}{\partial \bar \phi^i}$. 
From 
this K\"ahler potential (\ref{eq:kah}), 
the effective Lagrangian can be obtained as 
\beq
\mathcal{L}^{eff} 
= \p_i \p_{\bar j} K(\phi,\bar \phi) \, \partial^\mu \phi^i \partial_\mu \bar \phi^j 
= K_{i\bar j}(\phi,\bar \phi) \, \partial^\mu \phi^i 
\partial_\mu \bar \phi^j.
\label{eq:general}
\eeq

\begin{figure}[htb]
\begin{center}
\includegraphics[height=4cm]{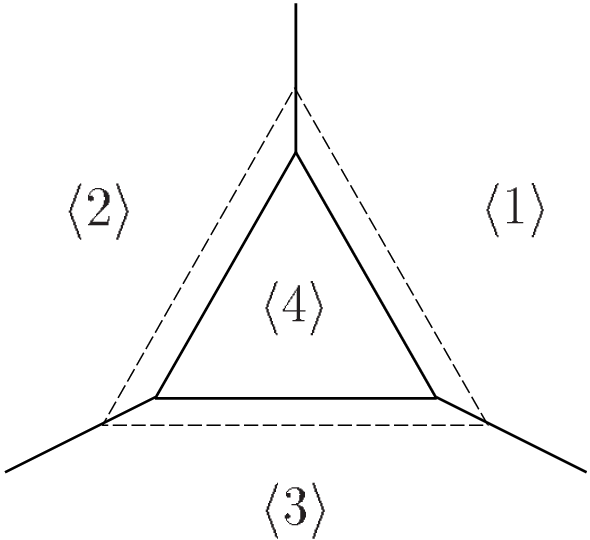}
\caption{\small 
The triangle loop configuration has four vacuum regions 
$\langle A \rangle$ $(A=1,\cdots,4)$.
We fix three complex moduli parameters which are related to 
positions of external walls. 
A unique normalizable mode is a zero mode which is related 
to the area of the region $\langle 4 \rangle$, namely the 
size of the triangle loop.}\label{fig:tri-mod}
\end{center}
\end{figure}

The domain wall network in Fig.\,\ref{fig:ab-grid}-(a) 
$(\NC=1,\ \NF=4)$ has 
a single normalizable complex zero mode. 
To describe the zero mode, we can take the following moduli 
matrix without loss of generality\footnote{
If we choose 
the V-transformation, the central position of 
the loop, and 
two relative phases carried by external walls, 
we can always reduce the moduli matrix $H_0$ for the single 
triangle loop to the form (\ref{eq:h_1_loop}).  
}
\beq
H_0 = \left(1,\ 1,\ 1,\ \phi\right),\qquad
\text{with} \quad
\phi = e^{w} = e^{r+i\theta}.
\label{eq:h_1_loop}
\eeq
The complex parameter $\phi$ is the normalizable modulus parameter.
One can easily see by looking at the weight of this system that 
its real part $r$ 
changes the configuration of the triangle loop as shown in 
Fig.\,\ref{fig:tri-mod}, and the imaginary part 
$\theta$ corresponds to the phase of the loop. 
For sufficiently large $r$,
the size of the loop is proportional to $r$.
 
The other zero modes, 
first three elements in Eq.~(\ref{eq:h_1_loop}),
are non-normalizable, 
and have to be fixed by boundary conditions 
when we construct the effective theory of the domain wall network. 
The effective Lagrangian 
for the size moduli have been already constructed in 
\cite{Eto:2006bb}. 
The K\"ahler potential and the metric are smooth everywhere 
in term of the complex coordinate $\phi$, and the moduli 
space has a geometry between a cone and a cigar with a tip 
at $\phi=0\,(r=-\infty)$. 
It was found that the K\"ahler potential 
in the
strong gauge coupling limit $g^2 \to \infty$ 
is given as a sum of hypergeometric functions, 
see Eq.~(3.30) of 
\cite{Eto:2006bb}.

In particular the K\"ahler potential and 
the metric of the single triangle loop 
are given asymptotically for large $|\phi|=e^r$ by 
\beq
K &=& \frac{c}{4\Delta_{[123]}}
\left[ \frac{1}{6\alpha_1\alpha_2\alpha_3}
\left(\log|\phi|^2\right)^3
\mp \frac{1}{g^2c} 
\left(\frac{|\vec m_{12}|^2}{\alpha_3}+ 
\frac{|\vec m_{23}|^2}{\alpha_1}
+\frac{|\vec m_{31}|^2}{\alpha_2}\right)
\left(\log|\phi|^2\right)^2\right],
\label{eq:tri-asym1}
\\
ds^2 &=&
\frac{c}{\Delta_{[123]}}
\left[ \frac{r}{\alpha_1 \alpha_2 \alpha_3}
\mp \frac{1}{g^2 c}
\left(\frac{|\vec m_{12}|^2}{\alpha_3}
+\frac{|\vec m_{23}|^2}{\alpha_1}
+\frac{|\vec m_{31}|^2}{\alpha_2}\right)\right](dr^2+d\theta^2),
\label{eq:tri-asym}
\eeq
where $\vec m_{AB} = \vec m_B - \vec m_A$ and 
ratios $\alpha_A \equiv \frac{1}{2\Delta_{[123]}} 
\epsilon_{ABC} \, \vec m_B \times \vec m_C$ 
satisfying $\alpha_1+\alpha_2+\alpha_3=1$. 
The minus sign in Eqs.(\ref{eq:tri-asym1}), (\ref{eq:tri-asym}) 
is for the triangle loop in the $U(1)$ gauge theory 
and plus sign in the $U(3)$ gauge theory\footnote{
Here we take 
gauge coupling $g$ and mass parameters for Higgs scalars 
(hypermultiplets) to be the same 
for two distinct $U(1)$ and $U(3)$ gauge theories with 
$N_{\rm F}=4$. 
They are dual in the sense that the number of vacua is equal and also their 
grid diagrams are congruent to each other. 
The duality becomes exact 
in the strong gauge coupling limit $g^2 \to \infty$. 
}. 
The corrections to the asymptotic metric 
have been found to be exponentially suppressed. 
An interesting feature is that 
the above asymptotic metric 
can be understood as the kinetic energies: 
the first terms in the parentheses in 
Eqs.(\ref{eq:tri-asym1}), (\ref{eq:tri-asym}) represent 
the kinetic energies of domain walls 
and the second (with the $\mp$ sign in front) 
that of junctions.
Since the lengths of domain walls are proportional to $r$, 
their masses and kinetic energies have linear dependence on $r$.
On the other hand, the junction charges are localized at the junction points, and so
their kinetic energies do not depend on $r$.
This interpretation nicely 
explains 
the sign of the second term.
See Eq.\,(\ref{eq:abelian}) 
and Eq.\,(\ref{eq:non-abelian}). 
This result implies that the asymptotic metric 
for more complicated configurations 
can be also obtained by computing 
the kinetic energies 
of domain walls and junctions. 
We will often use this result 
in investigating the dynamics of 
various domain wall networks below. 

\begin{figure}[htb]
\begin{center}
\begin{tabular}{cc}
\includegraphics[width=60mm]{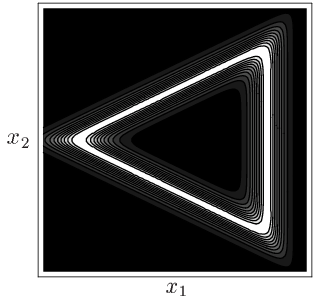} ~~~~&
\includegraphics[width=60mm]{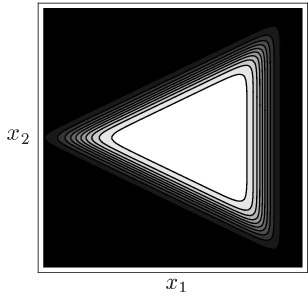}\\
~~(a) size and phase moduli & ~~~~~(b) vacuum moduli inside the loop
\end{tabular}
\end{center}
\caption{\small 
The densities of the metric for $\NC=1, \NF=5$ in strong coupling limit $g\rightarrow \infty$. At each point on the moduli space, the tangent space of the moduli space is orthogonally decomposed into the directions of size, phase and two vacuum moduli inside the loop.
}
\label{fig:intra}
\end{figure}

Before closing this section, let us briefly discuss
another configuration closely related to the above one.
When the vacua inside a loop are 
degenerate vacua, the loop acquires some internal moduli.\footnote{
Domain walls with degenerate masses were studied in 
\cite{Shifman:2003uh,Eto:2005cc}. 
It was found that some Nambu-Goldstone modes 
for broken non-Abelian flavor symmetry 
are localized around (between) the domain walls 
and appear in the effective theory on them. 
}
The asymptotic metric for the additional moduli
exhibits another characteristic feature.
Since the triangle loop requires at least 4 flavors that 
are non-degenerate, we need to take more than 4 flavors 
to examine a degenerate vacuum in the loop. 
Let us assume that mass parameters for external vacua are 
all non-degenerate $\vec m_A \neq \vec m_B$ for $A \neq B$ 
($A,B=1,2,3,4$), and the other mass parameters 
corresponding to the vacuum in the loop are all degenerate 
$\vec m_A = \vec m_B$ for ($4 \le A,B \le N_{\rm F}$). 
Such a loop configuration with the degenerate vacuum 
is described by the moduli matrix
\beq
H_0 = \left(1,\ 1,\ 1,\ 
{\boldsymbol \phi} \right),\quad
{\boldsymbol \phi} = \left(\phi_1,\phi_2,\cdots,
\phi_{\NF-3}\right).
\label{eq:moduli_matrix_degenerate}
\eeq
In this case, there exist $\NF-3$ complex normalizable zero modes: 
one of them corresponds to the size and phase of the loop 
and the others are zero modes associated with the vacuum moduli 
inside the loop.
We can obtain the K\"ahler potential $K$ in this case, if 
we replace $|\phi|^2$ in Eq.(\ref{eq:tri-asym1}) by 
$|{\boldsymbol \phi}|^2 \equiv |\phi_1|^2+\cdots
+|\phi_{N_{\rm F}-3}|^2$.
Therefore the knowledge of the K\"ahler potential for the 
$N_{\rm F}=4$ case gives a K\"ahler metric for this 
degenerate cases as 
\beq
K_{i\bar j} = \partial_{\phi_i}\partial_{\phi_{\bar j}} K 
= \delta_{ij}K'(|{\boldsymbol \phi}|^2) 
+ \bar \phi_i \phi_j K''(|{\boldsymbol \phi}|^2), 
\label{eq:kahler_metric_degenerate}
\eeq
where prime on $K$ denotes differentiation with respect to 
$|{\boldsymbol \phi}|^2$. 
 See Appendix \ref{appendix:degenerate} for a concrete example. 
By differentiating the leading contribution 
$(\log |{\boldsymbol \phi}|^2)^3$ of this 
K\"ahler potential at asymptotic region $r = \log |{\boldsymbol \phi}|^2 \gg 1$, 
we find that the K\"ahler metric 
contains not only terms proportional to $r$ as in 
Eq.(\ref{eq:tri-asym}), 
but also terms proportional to $r^2$
as shown in Eq.(\ref{eq:norm3}). 
This feature shows that among moduli fields, there are 
massless modes with a support extending two-dimensionally 
over the entire vacuum region inside the loop in the web 
of walls as illustrated in Fig.\,\ref{fig:intra}.

\section{Dynamics of Triangle Loop}\label{section:triangle}

\begin{figure}[htb]
\begin{center}
\includegraphics[width=60mm]{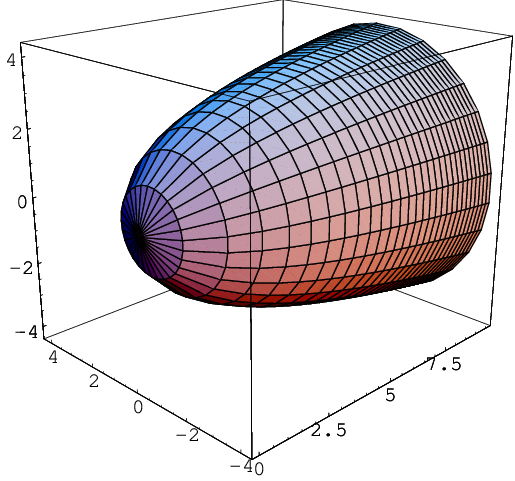}
\end{center}
\caption{\small The moduli space of single triangle loop embedded 
in $\mathbf R^3$: 
The moduli space has a $U(1)$ isometry which corresponds 
to the direction of the phase modulus. 
The other direction 
can be regarded as the direction of size modulus of the loop. 
The tip of the moduli space corresponds to the point 
$\phi=0$ where the loop shrinks completely.}
\label{fig:embed-tri}
\end{figure}

\begin{figure}[h]
\begin{center}
\includegraphics[width=70mm]{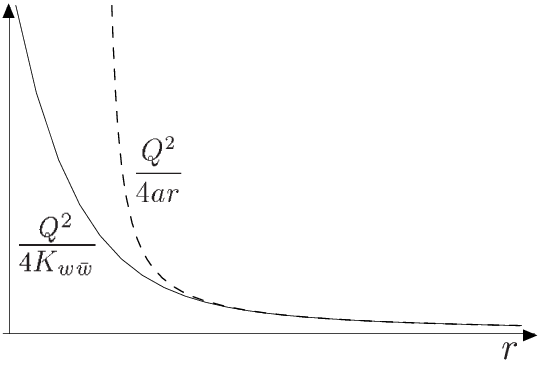}
\caption{\small Plot of the potential $V=Q^2/4K_{w \bar w}$ 
in $g^2 \rightarrow \infty$ limit (solid line) 
and asymptotic 
($r \gg 1$) potential $V=Q^2/4ar$, 
$a=\frac{c}{2\Delta_{[123]}}\frac{1}{\alpha_1 \alpha_2 \alpha_3}$ 
(dashed line) given in Eq.(\ref{eq:tri-asym}). 
The phase rotation produces 
the repulsive force among the triangle loop.}
\label{fig:potential}
\end{center}
\end{figure}

Since we have obtained the metric 
on the moduli space of the triangle loop, 
its dynamics can be discussed 
as geodesic motions on the moduli space. 
In order to avoid infinite volume of domain walls 
we may compactify the world-volume direction or 
simply dimensionally reduce the model to 
1+2 dimensions. Such model is obtained merely restricting the indices $M,N$
in the Lagrangian (\ref{eq:lag}) to $0,1,2$.
The effective Lagrangian takes the form 
\beq
L = K_{w \bar w}(r) \left[ \left(\frac{dr}{dt}\right)^2 
+ \left(\frac{d\theta}{dt}\right)^2 \right],
\label{eq:L_eff}
\eeq
with $w = r + i \theta$ 
in Eq.(\ref{eq:h_1_loop}). 
It is worth emphasizing that 
the moduli space is regular 
with positive curvature even when the loop shrinks completely. 
Fig.\,\ref{fig:embed-tri} shows 
the embedding of the moduli space into 3-dimensional Euclidean space. 
The moduli space has a $U(1)$ isometry 
which originates from a linear combination of 
three $U(1)$ flavor symmetries. 
Correspondingly, 
there exists a conserved charge such that 
\beq
Q \equiv \frac{\p L}{\p  (d\theta/dt)} = 2 K_{w \bar w} \frac{d\theta}{dt}.
\label{eq:Q-charge}
\eeq
In terms of this conserved $Q$-charge, 
the effective Lagrangian can be rewritten\footnote{ 
We have performed a Legendre transformation of $L$ with 
respect to $\theta$. 
}
as
\beq
\widetilde L 
= K_{w \bar w}(r) \, \left(\frac{dr}{dt}\right)^2 
- \frac{Q^2}{4 K_{w \bar w}(r)}.
\label{eq:L_eff2}
\eeq
Here the second term can be interpreted 
as a potential associated with the conserved charge $Q$. 
Note that the smoothness of the K\"ahler metric in terms of 
$\phi=e^w$ means the metric $K_{w\bar w}$ is exponentially 
suppressed as $K_{w\bar w}\propto e^{2r}\rightarrow 0$ 
for $r={\rm Re}\,w\rightarrow-\infty$. 
We also know the asymptotic behavior 
$K_{w\bar w}\propto r \rightarrow \infty$ for 
$r\rightarrow \infty$. 
The typical form of the potential is shown 
in Fig.\,\ref{fig:potential}. 
We can find that the phase rotation produces 
the repulsive potential among the triangle loop. 
This repulsive potential makes the loop to expand 
forever, namely the trajectory of the loop exhibits 
a runaway behavior.
Although we are now considering the effective theory of
domain wall networks, it is possible to consider domain wall networks
with $Q$-charges in the original theory. 
The above runaway potential tells us that such configuration is
 unstable and no longer BPS. However,
we will see in section \ref{sc:stabilization}, 
that a stable stationary point appears
if we introduce
another type of mass term (triplet mass). 
The corresponding configuration
will turn out to be BPS
, conserving a quarter of supercharges.

Now let us return to geodesic motions on the moduli space.
We introduce an 
integral of motion $E$ as an integration constant as 
\begin{equation}
E = K_{w \bar w}(r) \, \left(\frac{dr}{dt}\right)^2 
+ \frac{Q^2}{4 K_{w \bar w}(r)}, 
\end{equation}
corresponding to 
the energy associated with the motion of the zero-modes 
$r(t)$ and $\theta(t)$. 
By exploiting this conservation law of energy, 
we can obtain the solution of the equation of motion. 
The orbit of the geodesic for a given energy $E$ is given by 
\beq
\theta-\theta_0 = \pm 
\int dr \frac{Q}{\sqrt{4 K_{w \bar w}(r) E - Q^2}}, 
\eeq
and the time dependence of the size modulus $r$ 
is given by
\beq
t-t_0 = \pm 
 \int dr \frac{2 K_{w \bar w}(r)}{\sqrt{4 K_{w \bar w}(r) E - Q^2}}.
\eeq
If we consider the motion in the direction of smaller values 
of $r$ 
with $Q=0$, the geodesic is a straight line in the complex $\phi$-plane 
and goes through the tip of the manifold $\phi=0 \,\, (r=-\infty)$. 
This motion corresponds to the bounce of the loop, 
that is, after the loop shrinks completely, 
it tends to be larger with $180^\circ$ phase rotation. 
In the case of $Q \not = 0$, 
the repulsive force among the loop 
become stronger as the size become smaller, 
so that it prevents the loop 
from shrinking completely. 
Hence, there exists a minimum value of 
the size modulus $r$ determined by 
\beq
E= \frac{Q^2}{4K_{w \bar w}(r_{\rm min})}. 
\eeq
This implies 
that if the initial velocity of the size modulus $\frac{dr}{dt}$ is negative, 
the loop shrinks to its minimum size $r=r_{\rm min}$ 
and then the velocity $\frac{dr}{dt}$ changes its sign. 

\begin{figure}[htb]
\begin{center}
\includegraphics[height=4cm]{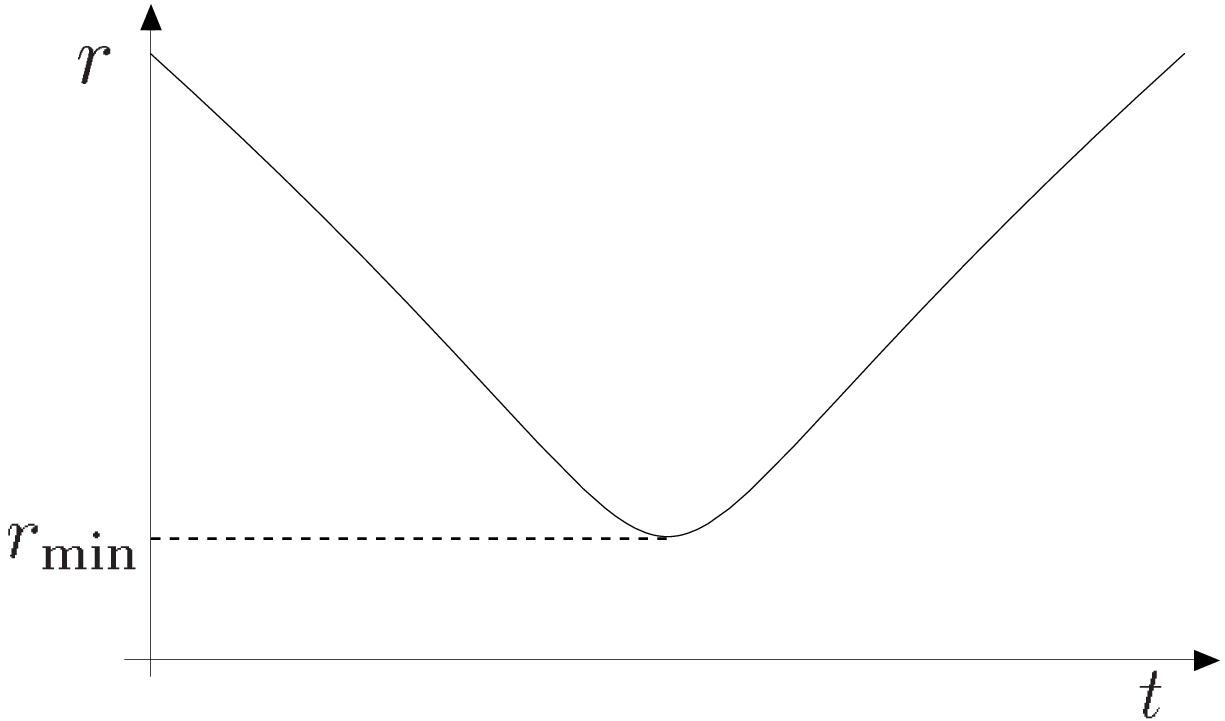}
\caption{\small The solution of the equation of motion for size modulus with $Q \not = 0$. The phase rotation produces the repulsive potential 
and the loop bounces back at $r_{\rm min} = Q^2/4aE$.}
\label{fig:bounce}
\end{center}
\end{figure}

The large size behavior can be investigated 
using the asymptotic metric Eq.\,(\ref{eq:tri-asym}). 
First note that the second term in Eq.\,(\ref{eq:tri-asym}) 
can be absorbed by shifting the parameter 
as $r \rightarrow r \mp \alpha_1 \alpha_2 \alpha_3 
\left( |\vec m_{12}|^2/\alpha_3 + |\vec m_{23}|^2/\alpha_1 
+ |\vec m_{31}|^2/\alpha_2 \right)/g^2 c$. 
After the shift, 
the equation of motion for $r$ 
can be solved as 
\beq
\theta-\theta_0 &=& \pm \frac{Q}{2aE}\sqrt{4 a E r - Q^2}, \label{eq:bounce1}\\
t-t_0 &=& \pm \frac{1}{6 a E^2} \left( 2 a E r + Q^2 \right) \sqrt{4 a E r - Q^2},
\label{eq:bounce2}
\eeq
where $a = \frac{c}{2\Delta_{[123]}}\frac{1}{\alpha_1 \alpha_2 \alpha_3}$.
In the case of $Q=0$, 
the above equation says $r \sim t^{\frac{2}{3}}$. 
This reflects the fact that 
the mass of the triangle loop is proportional to $r$ 
and its velocity becomes smaller 
as the size of the loop becomes large.
For the loop with $Q \neq 0$, 
the minimum size is given by $r_{\rm min} = Q^2/4aE$.
The typical time dependence of the size modulus 
is shown in Fig.\,\ref{fig:bounce}.
Since our argument above is based on 
the asymptotic metric Eq.\,(\ref{eq:tri-asym}) 
which is valid for $r\gg1$, 
the solution Eq.\,(\ref{eq:bounce1}) 
and Eq.\,(\ref{eq:bounce2}) can be well trusted 
only when $r_{\rm min} = Q^2/4aE \gg 1$. 
Of course, the energy $E$ should be small enough 
so as not to excite the massive modes. 

The asymptotic potential $V=Q^2/4ar$ 
can be interpreted as the shift of the energies 
associated with the walls composing the loop. 
This expectation can be confirmed 
by the following argument. 
For a domain wall with tension $T$, 
the rotation of its phase 
induces flavor charge density $\rho_Q$ 
on the world-volume given by 
\beq
\rho_Q \,=\, \frac{c^2 \frac{d\theta}{dt}}{\sqrt{T^2 - c^2 \left(\frac{d\theta}{dt}\right)^2}} 
\,\sim\, \frac{c^2}{T}\frac{d\theta}{dt}.
\eeq
In addition, the phase rotation cause a shift of the tension as 
\beq
\Delta T \,=\, T \sqrt{1+\rho_Q^2/c^2} - T 
\,\sim\, \frac{T}{2c^2} \rho_Q^2.
\eeq
For the loop, each wall composing the loop 
becomes a domain wall with flavor charges 
by rotating the phase modulus $\theta$. 
Since the tensions and lengths of the walls 
composing the loop are given by
\beq
T^{(A,4)} = c |\vec{m}_A|, \hs{5} 
l^{(A,4)} = \frac{|\vec{m}_A|}{\Delta_{[123]}} \frac{\alpha_A}{\alpha_1\alpha_2\alpha_3} r, \hs{10} A=1,2,3, 
\label{eq:tention-length}
\eeq
the total flavor charge for the wall 
interpolating between $\langle A \rangle$-th 
and $\langle 4 \rangle$-th vacua is
\beq
Q_A \,=\, \frac{c^2}{T^{(A,4)}}\frac{d\theta}{dt}\,\,l^{(A,4)} 
\,=\, \frac{c}{\Delta_{[123]}} \frac{\alpha_A}{\alpha_1\alpha_2\alpha_3} r\frac{d\theta}{dt}.
\eeq
From this expression, 
we find that the total charge $Q = \sum Q_A$ agrees 
with Eq.\,(\ref{eq:Q-charge}) in large $r$ limit 
and $Q_A$ are given by $Q_A = \alpha_A Q$. 
Therefore the total shift of energy is given by
\beq
\sum_{A=1}^3 \Delta T^{(A,4)} l^{(A,4)} 
\,=\, \sum_{A=1}^3 \frac{T^{(A,4)}}{2c^2}\left( \frac{\alpha_A Q}{l^{(A,4)}} \right)^2 l^{(A,4)} 
\,=\, \frac{Q^2}{4ar}.
\eeq
From the argument above, 
we can intuitively understand the reason 
why the phase rotation give rise to 
a repulsive potential as follows: 
Since the shifts of tensions of the walls 
are proportional to $\rho_Q^2$ 
and the lengths of the walls are 
proportional to $r$, 
the shift of the total energy is 
proportional to $\rho_Q^2 r$. 
This fact implies that 
if the charge density is constant, 
the total energy increases 
as the size of the loop becomes larger. 
However since the conserved quantity is 
not the density $\rho_Q$ 
but the total flavor charge $Q$, 
the density decreases in proportion to $1/r$ 
and hence the total energy decreases 
in proportion to $1/r$ 
as the size of the loop become larger. 
This is the reason why 
the phase rotation produces the 
decreasing repulsive potential. 

Next, let us consider the dynamics of 
a triangle loop with degenerate masses.
The K\"ahler potential can be obtained by replacing $|\phi|^2$ with $|\boldsymbol \phi|^2 \equiv |\phi_1|^2 + \cdots |\phi_{\NF-3}|^2$ in 
$K(|\phi|^2)$ and the K\"ahler metric of the moduli space is given in Eq.\,(\ref{eq:kahler_metric_degenerate}). All but the size moduli $r \equiv \log |\boldsymbol \phi|$ can be eliminated from the expression of energy by using the conserved charges associated with $U(\NF-3)$ flavor symmetry. For example, the energy in the case of $\NF=5$ takes the form (See Appendix \ref{appendix:degenerate})
\beq
E = \frac{1}{4} \p_r^2 K \, \dot r^2 + 
\frac{Q^2}{\p_r^2 K}+\frac{|q|^2-Q^2}{2\p_r K}, 
\label{eq:degenergy}
\eeq
where $|q|^2\equiv \sum_{a=1}^3q_aq_a$ and the conserved 
charges $Q,\,q_a\,(a=1,2,3)$ are defined by 
\beq
Q = i K_{i \bar j} \left( \frac{d \bar \phi^j}{dt} \phi^i 
- \frac{d\phi^i}{dt} \bar \phi^j \right), \hs{10} 
q_a = i K_{i \bar j} \left( \frac{d \bar \phi^j}{dt} 
{\left( \sigma_a \right)^i}_k \phi^k 
- \frac{d\phi^i}{dt} \bar \phi^k 
{\left( \sigma_a \right)_k}^j \right).
\eeq
Note that these conserved charges are related as 
$Q = (\bar \phi \sigma_a \phi) q_a/|\boldsymbol \phi|^2$ 
and satisfy an inequality  $Q^2\le|q|^2$. 
The second and third terms in Eq.\,(\ref{eq:degenergy}) can be interpreted as the effective potential $V(r)$ associated with the conserved charges. For large $r$, this potential takes the form
\beq
V(r) = \frac{Q^2}{4ar} + \frac{|q|^2-Q^2}{4ar^2}
\label{eq:r-2potential}
\eeq
The first term of the potential takes the same form as in 
the case of non-degenerate masses. 
Conversely, the second term is induced by the Noether 
charges associated with the vacuum moduli inside the loop. 
To understand intuitively the origin of the second term, 
we can use the same argument for the first term. 
What we should notice is only the fact that 
a part of the Noether charge $q_a$ which have no 
contribution to $Q$ is supported by the 
two-dimensional vacuum inside the loop, while the charge 
$Q$ has one-dimensional support on the walls.
Therefore we can easily re-derive the behavior of the 
second term repulsive potential proportional to $1/r^2$. 
This potential can be also understood by exchange of 
the massless particles propagating the degenerate vacuum. 
\section{Dynamics of Double Loop}\label{section:double}

We will consider the dynamics of double loop 
shown in Fig.\,\,\ref{fig:ab-grid}-(b) in this section. 
Unlike the previous example, 
this configuration has two normalizable zero modes 
which are related to the sizes of the loops and their phases. 
By varying the sizes of the loops, 
we obtain various configurations of the loops. 
We will first explain the configurations of the domain wall web 
and then discuss the dynamics of the double loop.

\begin{figure}[htb]
\begin{center}
\begin{minipage}{8.5cm}
\vspace{0.5cm}
\begin{center}
\includegraphics[width=7.5cm]{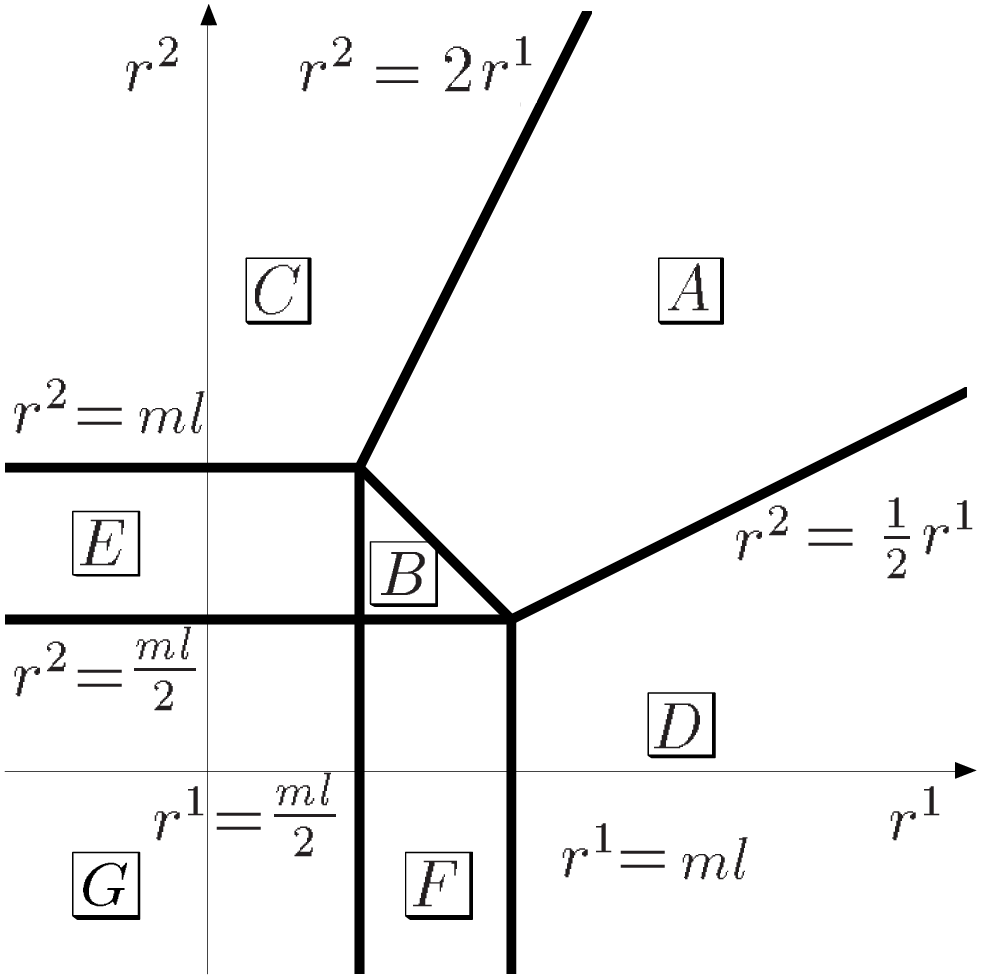}
\end{center}
\vspace{-0.7cm}
\caption{\small Seven regions of moduli space 
corresponding to different patterns of 
configurations. 
}
\label{fig:configdiagram}
\end{minipage}
\begin{minipage}{8.5cm}
\begin{center}
\vspace{-0.9cm}
\includegraphics[width=8.5cm]{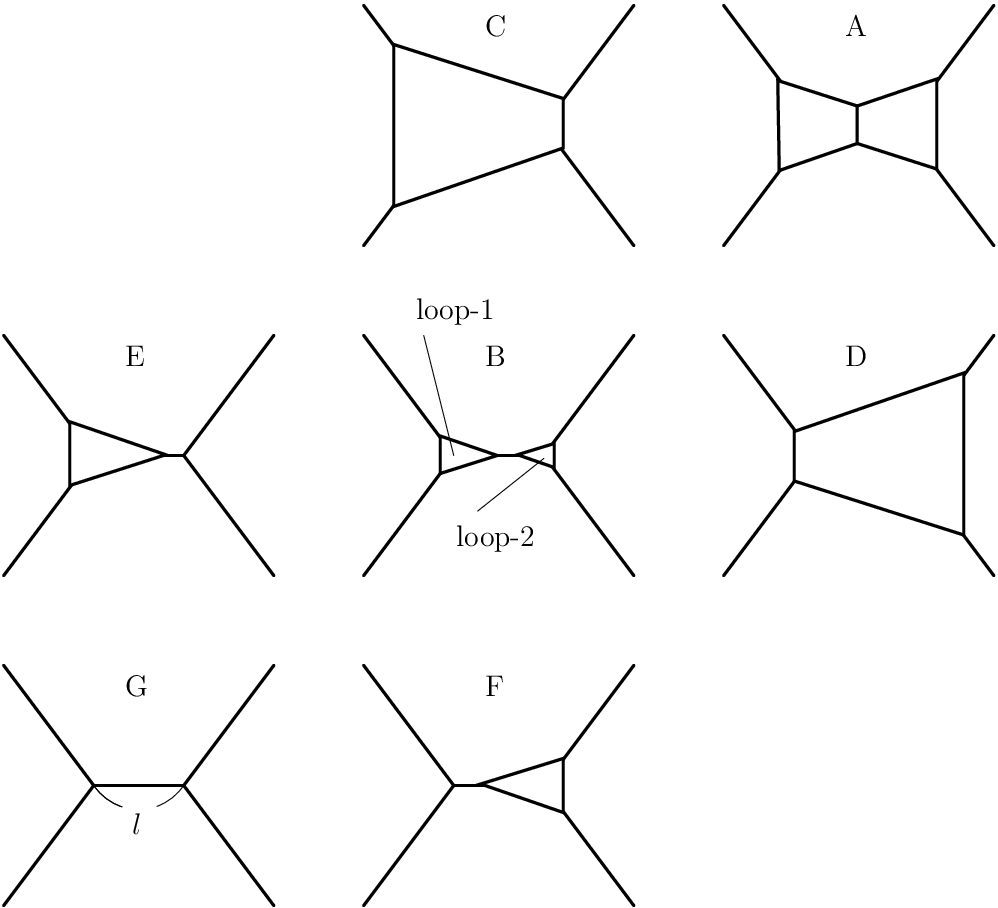}\vspace{-0.4cm}
\end{center}
\caption{\small Configurations of double loop.}
\label{fig:configurations}
\end{minipage}
\end{center}
\end{figure}

The model is $U(1)$ gauge theory with six hypermultiplets 
and we choose six complex masses as follows 
(assuming a real positive value for $m>0$): 
\beq
M={\rm diag.}\,\left(\frac{3m}{2},\,i\frac{\sqrt{3}m}{2},
\,-\frac{3m}{2},\,-i\frac{\sqrt{3}m}{2},
\,\frac{m}{2},\,-\frac{m}{2} \right).
\label{eq:mass_double_loop}
\eeq
The solution of the BPS equations 
Eq.\,(\ref{eq:BPS2}) and Eq.\,(\ref{eq:BPS1}) 
is characterized by $H_0$, 
which is now six component row vector 
\beq
H_0=\sqrt{c}\,(e^{a_1+ib_1}, e^{a_2+ib_2}, e^{a_3+ib_3}, e^{a_4+ib_4},
e^{a_5+ib_5}, e^{a_6+ib_6}).
\eeq
Since the parameters $a_i$ and $b_i$ $(i=1,\cdots,4)$ 
are related to positions and phases of external walls, 
these will be non-normalizable modes 
if these are promoted to fields, 
and normalizable zero modes correspond to 
the parameters in the fifth and sixth components of $H_0$. 
They are related to sizes of the double loop and their phases. 
When we consider the effective theory of the domain wall network, 
we have to fix four complex moduli parameters 
and promote two normalizable modes to fields: 
\beq
H_0=\sqrt{c}\,(1,\,e^{3ml/4},\,1,\,e^{3ml/4},\,\phi^1(x^\mu),\,\phi^2(x^\mu))
\eeq
with $\phi^i = e^{w^i} = e^{r^i+i\theta^i}~(i=1,2)$.
For simplicity, we have chosen a somewhat symmetric set of 
four complex parameters for external walls. 
By varying the sizes of loops, we obtain seven 
different patterns of web configurations as shown 
in Fig.\,\ref{fig:configdiagram} and 
Fig.\,\ref{fig:configurations}. 
Let us recall the vacuum assignment depicted in 
Fig.\ref{fig:ab-grid}(b) which gives the grid diagram 
together with the web diagram A in 
Fig.\ref{fig:configurations}. 
The fixed parameter $l$ corresponds to the length between 
$\langle 1 \rangle \langle 2 \rangle\langle 4 \rangle$ junction
and $\langle 2 \rangle \langle 3 \rangle\langle 4 \rangle$ junction,
in other words, the length of $\langle 2 \rangle \langle 4 \rangle$ wall 
(internal line) of configuration G in Fig.\,\ref{fig:configurations}.
We call the left loop surrounding vacuum $\left<5\right>$ loop-1 and
the right surrounding $\left<6\right>$ loop-2. 
In the region A(B), both
the loop-1 and the loop-2 appear as quadrangle loops(triangle loops).
When the loop-1(2) grows and covers the junction 
$\langle 1 \rangle \langle 2 \rangle\langle 4 \rangle$
($\langle 2 \rangle \langle 3 \rangle\langle 4 \rangle$), 
the other loop-2(1) is eaten by the loop-1(2) as C(D) in Fig.~\ref{fig:configurations}.
In the region E(F) the loop-2(1) vanishes and the triangle loop-1(2) exists.
In the region G both the loop-1 and loop-2 disappear.

Since the K\"ahler potential $K$ is independent of $\theta^i$, 
the K\"ahler metric $K_{i \bar j}$ can be written as 
$K_{i \bar j} \equiv \frac{\p}{\p w^i} \frac{\p}{\p \overline w^j} K = \frac{1}{4} \frac{\p}{\p r^i} \frac{\p}{\p r^j} K$. 
Then the effective Lagrangian takes the form 
\beq
L = K_{i \bar j}(r^1,r^2) \frac{d{w}^i}{dt} \frac{d{\overline{w}}^j}{dt} =
K_{i \bar j}(r^1,r^2) \left( \frac{dr^i}{dt} \frac{dr^j}{dt} 
+ \frac{d\theta^i}{dt} \frac{d\theta^j}{dt} \right),
\eeq
where we have used $K_{i \bar j} = K_{j \bar i}$. 
In this case, there exist two conserved charges defined by 
$Q_i \equiv 2 K_{i \bar j} \frac{d\theta^j}{dt}$. 
By using these conserved 
charges, 
the Lagrangian can be rewritten as 
\beq
\widetilde{L} = K_{i \bar j}(r^1,r^2) 
\frac{dr^i}{dt} \frac{dr^j}{dt} - \frac{1}{4} K^{\bar j i}(r^1,r^2) Q_i Q_j,
\eeq
where $K^{\bar j i}$ is the inverse of the metric $K_{i \bar j}$. 

\begin{figure}[h]
\begin{center}
\begin{tabular}{ccc}
\includegraphics[height=52mm]{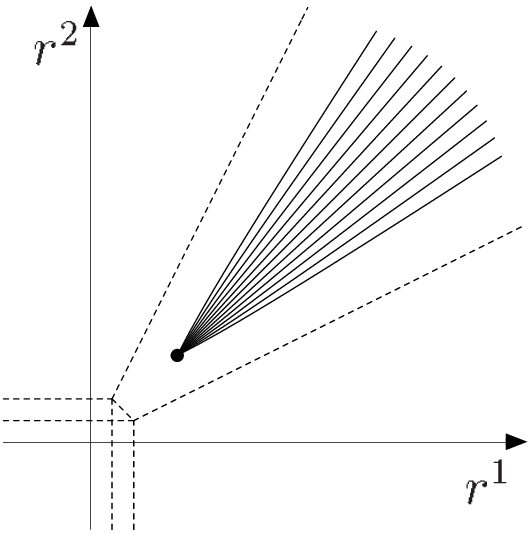}&
\includegraphics[height=52mm]{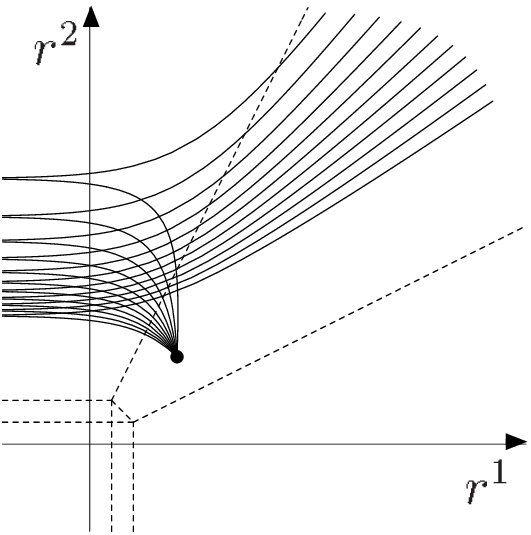}&
\includegraphics[height=52mm]{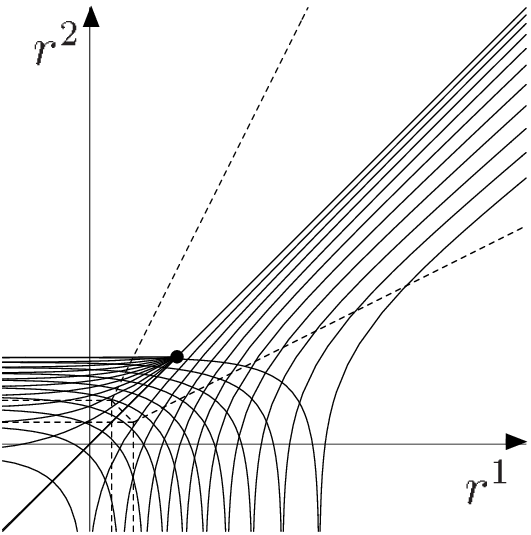}\\
(a) $\frac{dr^2}{dt} \lesssim 2\frac{dr^1}{dt}$, $\frac{dr^2}{dt} \gtrsim 
\frac{1}{2} \frac{dr^1}{dt}$&(b) $\frac{dr^2}{dt} \gtrsim 2 \frac{dr^1}{dt}$, 
$\frac{dr^2}{dt} \gtrsim -\frac{dr^1}{dt}$&(c) $\frac{dr^2}{dt} 
\lesssim -\frac{dr^1}{dt}$, $\frac{dr^2}{dt} \gtrsim \frac{1}{2} \frac{dr^1}{dt}$
\end{tabular}
\caption{\small Numerical solutions of the equation of motion for 
double loop 
without flavor charges. The initial state has taken to be configuration A 
with the same loop size and some orbits for various initial velocities 
are shown in these figures.}
\label{fig:sml}
\end{center}
\end{figure}

In the previous section, 
we have found the characteristic property of loop, 
that is, the loop is apt to become larger irrespective of 
$Q=0$ or $Q \neq 0$. 
Therefore, we expect that 
the loops would become larger 
and sit in region A in Fig.\,\ref{fig:configdiagram} 
after sufficiently long time interval. 
Some examples of numerical solutions 
without flavor charges ($Q_1=Q_2 = 0$) 
are shown in Fig.\,\ref{fig:sml}. 
For the initial velocities such that 
$\frac{dr^2}{dt} \lesssim 2 \frac{dr^1}{dt}$ 
and $\frac{dr^2}{dt} \gtrsim \frac{1}{2} \frac{dr^1}{dt}$ 
(Fig.\,\ref{fig:sml}-(a)), 
the orbits of the solutions are almost straight lines 
in $r^1$-$r^2$ plane and sit entirely in region A. 
For the initial velocities such that 
$\frac{dr^2}{dt} \gtrsim 2 \frac{dr^1}{dt}$ 
and $\frac{dr^2}{dt} \gtrsim -\frac{dr^1}{dt}$ (Fig.\,\ref{fig:sml}-(b)), 
the orbits of the solutions first enter region C, 
namely one of the loops shrinks. 
Then, they bounce back at $r^1=-\infty$ and return to region A. 
For the initial velocities such that 
$\frac{dr^2}{dt} \lesssim -\frac{dr^1}{dt}$ 
and $\frac{dr^2}{dt} \gtrsim \frac{1}{2} \frac{dr^1}{dt}$ (Fig.\,\ref{fig:sml}-(c)), 
one of the loops shrinks 
and bounces back at $r^1=-\infty$. 
Then, they enter region D and bounce at $r^2=-\infty$,
namely the other loop shrinks to zero size. 
Finally, the orbits return to region A 
and the sizes continue to become larger. 
\begin{figure}[h]
\begin{center}
\begin{tabular}{cc}
\includegraphics[height=6cm]{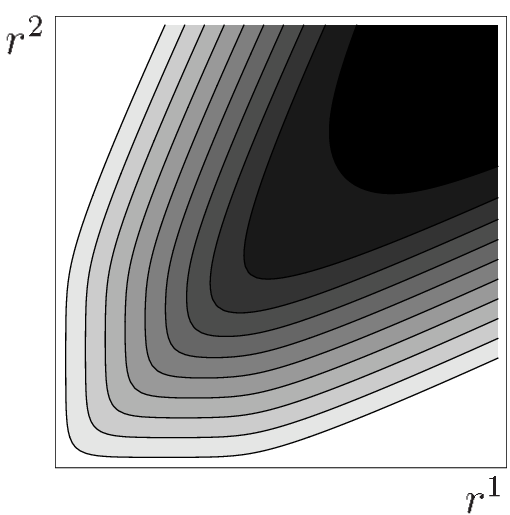}&
\includegraphics[height=6cm]{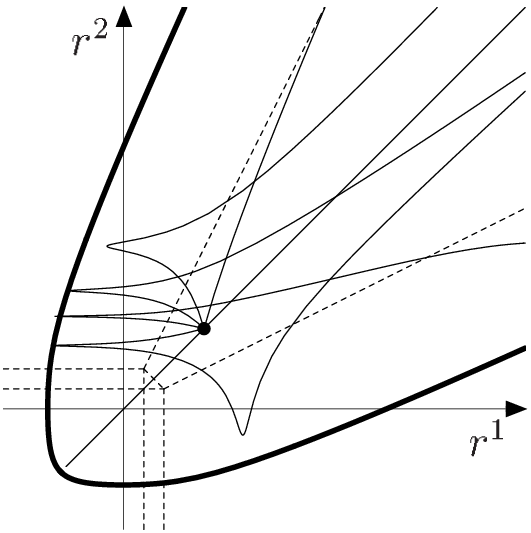} \\
(a) Contour plot of the potential, &
(b) numerical solutions with flavor charges. 
\end{tabular}
\caption{\small (a) Contour plot of the potential 
$\log V = \log \left( \frac{1}{4}K^{\bar j i} Q_i Q_j \right)$. 
(b) Orbits of numerical solutions with flavor charges. 
There exists a forbidden region where the potential energy exceeds the given total energy $V(r)>E$.}
\label{fig:dlQ}
\end{center}
\end{figure}

Next, let us consider the case of the double loop 
with the flavor charges. 
Fig.\,\ref{fig:dlQ} shows an example of the potential 
$V=\frac{1}{4}K^{\bar j i} Q_i Q_j$ 
and numerical solutions for the double loop with flavor charges. 
The potential increases rapidly outside region A 
and produces the repulsive force among the walls, 
so that any orbits of the solutions enter region A 
after sufficiently long time interval 
and continue to become larger. 
Other numerical simulations also demonstrate 
that region A is preferable. 

It is possible to know the asymptotic metric 
in region A
for $r^1 \approx r^2 \gg ml$ by computing 
the kinetic energy of domain walls. The kinetic energy of domain walls 
implies that the asymptotic metric $r^1 \approx r^2 \gg ml$ is given by
\beq
ds^2 &=& \frac{c}{\sqrt{3}m^2}
\bigg[(8r^1-r^2)|dw^1|^2+(8r^2-r^1)|dw^2|^2 - (r^1+r^2) (dw^1 d\bar{w}^2 + dw^2 d\bar{w}^1) \bigg] \notag \\
&=& \frac{5\sqrt{3}c}{2m^2}\bigg[
{\rm Re} \, \mu_+ \, |d\mu_+|^2 +
{\rm Re} \, \mu_- \, |d\mu_-|^2 \bigg],
\eeq
where $\mu_\pm$ is defined by $\mu_\pm \equiv (1 \pm \lambda)w^1/2 +
(1 \mp \lambda)w^2/2,~\lambda \equiv 3/\sqrt{5}$. 
This form of the metric implies that the geodesic equation decomposes into 
two independent equations which can be solved as in the case of the triangle loop. 

Fig.\ref{fig:dlQ} appears to illustrate that
the trajectory of the double loop 
configurations can bounce back at most only twice
. However, this behavior is due to the particular mass assignment 
of the model,
namely the center wall in configuration A is rather heavy.
Let us consider smaller mass difference between two flavors 
corresponding to the vacua inside the loop, such as 
\beq
M={\rm diag.}\,\left(\frac{3m}{2},\,i\frac{\sqrt{3}m}{2},
\,-\frac{3m}{2},\,-i\frac{\sqrt{3}m}{2},
\,\frac{m'}{2},\,-\frac{m'}{2} \right), \quad 
m' \ll m.
\label{eq:mass_double_loop1}
\eeq
Then the mass of the center wall is much smaller than those of the other walls. 
Such mass assignment makes it possible that
the double loop configuration bounces a lot of times. Moreover, 
if we consider the case of degenerate masses $m'=0$, 
we can have configurations exhibiting as many 
``bounces" as one wishes. 
In this degenerate mass limit, 
the center wall is no longer visible, rather, it spreads 
over the entire middle vacuum region in the loop. 
Correspondingly, the mode also spreads over the vacuum 
region as in the case of a triangle loop with degenerate 
masses (See Fig.\,\ref{fig:intra}) and it describes the 
degrees of freedom of the degenerate vacua inside the 
quadrangle loop. 
Furthermore, 
it is interesting to note that the infinitely 
many ``bounces'' in the degenerate mass limit naturally 
reduces to 
the repulsive force, which is described by a potential 
similar to the second term in Eq.(\ref{eq:r-2potential}) 
with a different coefficient $a$. 
To describe the dynamics in the degenerate mass limit, 
we can use the expression for the energy 
Eq.(\ref{eq:degenergy}), if the K\"ahler potential $K$ 
is replaced by that for a quadrangle loop.

\section{Dynamics of Non-Abelian Loop}\label{non-abe}

We will next consider the dynamics of non-Abelian loop 
shown in Fig.\,\,\ref{fig:nab-grid} in this section. 
This configuration has four external walls, 
and also four internal walls 
which 
divide six vacua and constitute a quadrangle loop. 
After fixing the positions of external walls, 
one complex moduli parameter is left. 
The difference from the Abelian loop is 
that the moduli parameter controls 
the areas of two vacuum regions.
First we will explain the configuration and moduli parameters, 
and then we will discuss the dynamics of the non-Abelian loop. 
For the details of the non-Abelian webs of walls, see \cite{Eto:2005fm}.

\begin{figure}[htb]
\begin{center}
\begin{tabular}{ccc}
\includegraphics[height=4cm]{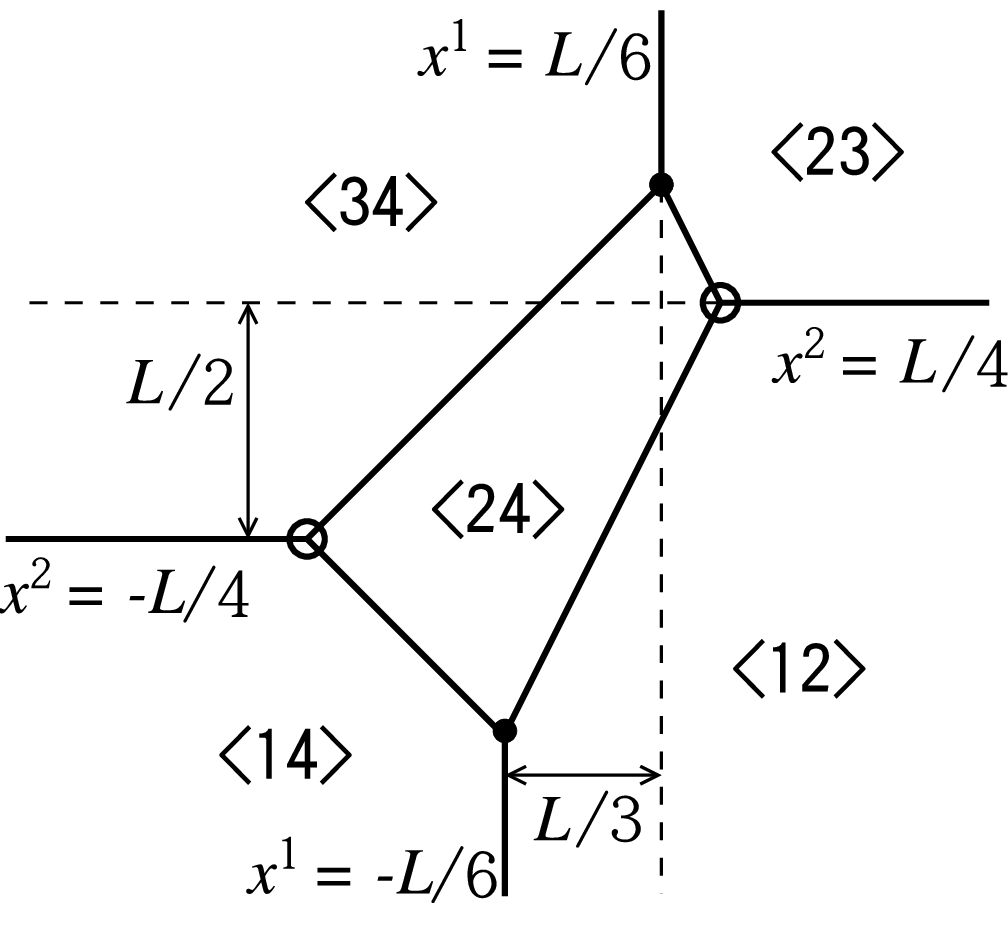}
&\qquad&
\includegraphics[height=4cm]{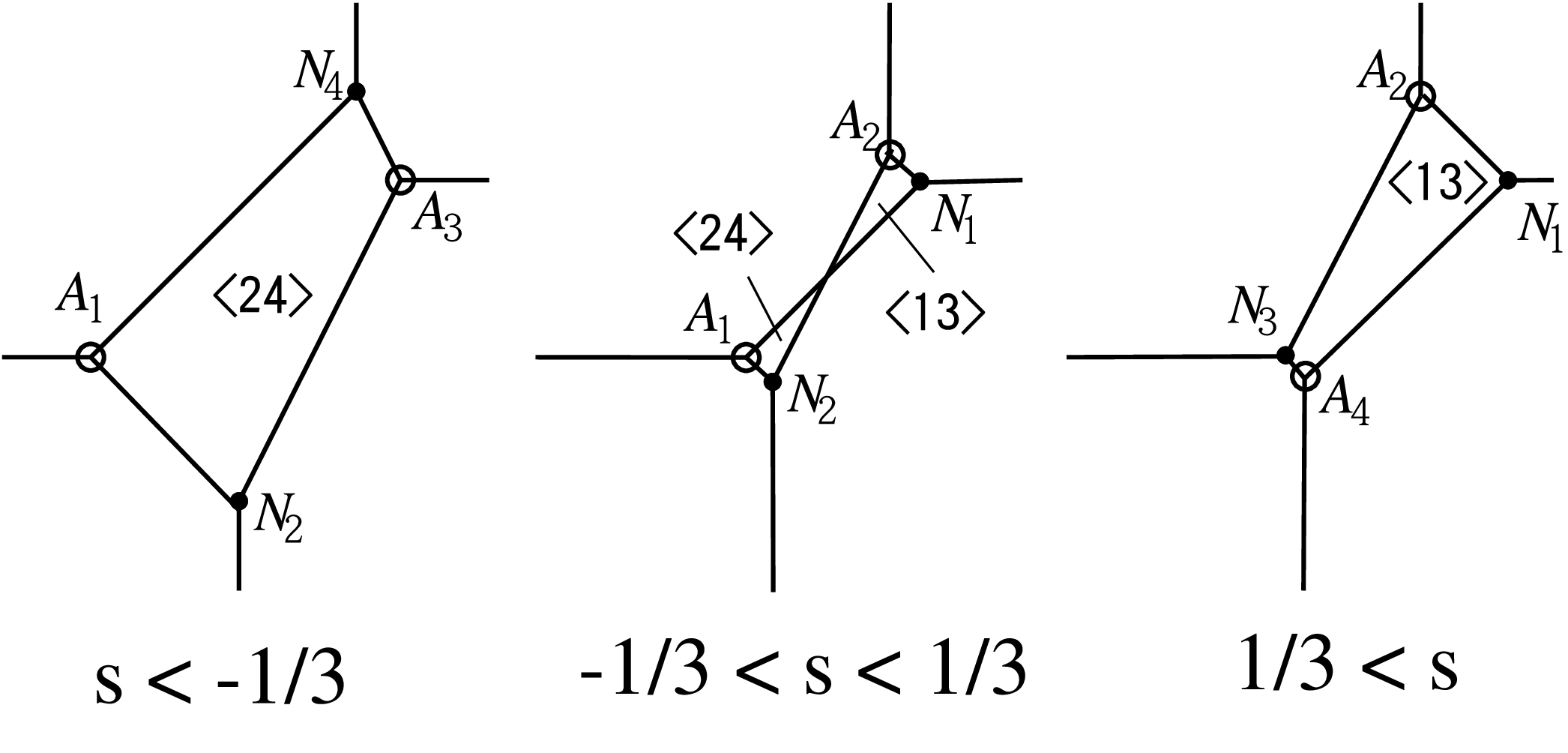}\\
\small{(a) A configuration} & & 
\small{(b) Deformation of the loop in the web}
\vspace*{-.3cm}
\end{tabular}
\caption{\small The web diagrams in the 
parallelogram-type mass arrangement.
Positions of the Abelian junction $A_*$ and 
the non-Abelian junction $N_*$ are given by
$A_1=(s-1,-1),
A_2 = (\frac{2}{3},\frac{4}{3}+s),
A_3 = (\frac{1-s}{2},1),
A_4= (-\frac{2}{3},-\frac{2}{3}-s),
N_1 = (1+s,1),
N_2= (-\frac{2}{3},s-\frac{4}{3}),
N_3= (-\frac{1+s}{2},-1),
N_4 = (\frac{2}{3},\frac{2}{3}-s)$ in unit of $\frac{L}{4}$.}
\label{fig:para}
\vspace*{-.3cm}
\end{center}
\end{figure}

The model is $U(2)$ gauge theory with $\NF=4$ hypermultiplets, 
and we choose four complex mass parameters as follows:
\beq
M_1+iM_2={\rm diag.}\,\left( \frac{m}{2}-im, \frac{3m}{2}, \frac{m}{2}+im, -\frac{3m}{2} \right).
\label{eq:quad_mass}
\eeq
The complex masses and vacuum points 
in the ${\rm Tr}\langle \Sigma \rangle$ plane are 
shown in Fig.\,\,\ref{fig:nab-grid}-(a).
The solutions of the BPS equations are 
characterized by $2 \times 4$ moduli matrix $H_0$. 
It is convenient to extract $2 \times 2$ matrix 
$H_0^{\langle A_1 A_2 \rangle}$ defined by 
$(H_0^{\langle A_1 A_2 \rangle})^{st}=(H_0)^{s A_t},~(s,t=1,2)$. 
Let us denote $\det H_0^{\langle A_1 A_2 \rangle}$ as
\beq
\tau^{\langle A_1 A_2 \rangle} \equiv \exp (a^{\langle A_1 A_2 \rangle}+ib^{\langle A_1 A_2 \rangle})
=\det H_0^{\langle A_1 A_2 \rangle}.
\label{eq:tau}
\eeq
These parameters are not independent but satisfy 
the so-called Pl\"ucker relation given by
\beq
\tau^{\langle 12 \rangle}\tau^{\langle 34 \rangle}-
\tau^{\langle 13 \rangle}\tau^{\langle 24 \rangle}+
\tau^{\langle 14 \rangle}\tau^{\langle 23 \rangle}=0.
\label{eq:plucker}
\eeq
Each parameter $a^{\langle A_1 A_2 \rangle}$ 
corresponds to the area of 
the vacuum region $\langle A_1 A_2 \rangle$ 
and $b^{\langle A_1 A_2 \rangle}$ 
to the associated phase as before. 
In order to fix four external walls, 
we set four complex moduli parameters as
\begin{align}
&a^{\langle 12 \rangle}=a^{\langle 34 \rangle}
=-a^{\langle 14 \rangle}=-a^{\langle 23 \rangle}=\frac{mL}{4},\\
&b^{\langle 12 \rangle}=b^{\langle 34 \rangle}=b^{\langle 14 \rangle}=b^{\langle 23 \rangle}=0.
\label{eq:quad_fix}
\end{align}
The parameter $L$ controls the positions of the external walls 
and the shape of the quadrangle loop 
as shown in Fig.\,\ref{fig:para}.
The remaining moduli parameters are
$\tau^{\langle 13 \rangle}$ and $\tau^{\langle 24 \rangle}$, which
determine the size of the loop. 
We introduce two complex parameters $u,v \in {\bf C}$ as
\begin{align}
a^{\langle 13 \rangle}+ib^{\langle 13 \rangle}=(u+v)\frac{mL}{4},\\
a^{\langle 24 \rangle}+ib^{\langle 24 \rangle}=(u-v)\frac{mL}{4}.
\end{align}
The parameter $u$ is fixed 
by the Pl\"ucker relation Eq.\,(\ref{eq:plucker}). 
In the following, we take $L$ sufficiently large, $L \gg 1/m$, 
so that the equation Eq.\,(\ref{eq:plucker}) 
determines the parameter $u$ as $u \simeq 1$. 
Then the only parameter left is $v$, 
which we denote as $v \equiv s + i\theta$. 
The moduli parameter $s$ controls the areas of two vacua
$\left<13\right>$ and $\left<24\right>$, 
and three patterns of webs with a quadrangle loop 
appear as $s$ changes,  
as shown in Fig.\,\ref{fig:para}. 
The parameter $\theta$ is related to 
the Nambu-Goldstone mode corresponding to 
one of the broken flavor symmetries.

Now let us discuss the dynamics of the non-Abelian loop. 
If we calculate the kinetic energies 
of domain walls separately 
in three configurations in Fig.\,\ref{fig:para}, 
the asymptotic metric on the moduli space 
of the quadrangle loop is obtained as 
\beq
\begin{array}{ccll}
ds_{\rm w}^2 &=& \displaystyle \frac{3mc}{2}\left(\frac{L}{4}\right)^3(s+1)(ds^2+d \theta^2),
&\hs{10} s \gg \displaystyle \frac{1}{3}, \\
ds_{\rm w}^2 &=& \displaystyle 2mc \left(\frac{L}{4}\right)^3(ds^2+d \theta^2),
&\hs{10} \displaystyle s \approx 0, \\
ds_{\rm w}^2 &=& \displaystyle \frac{3mc}{2}\left(\frac{L}{4}\right)^3(-s+1)(ds^2+d \theta^2),
&\hs{10} s \ll - \displaystyle \frac{1}{3}.
\end{array}
\label{quad-wall}
\eeq
\begin{figure}[htb]
\begin{center}
\begin{tabular}{cc}
\includegraphics[width=7cm]{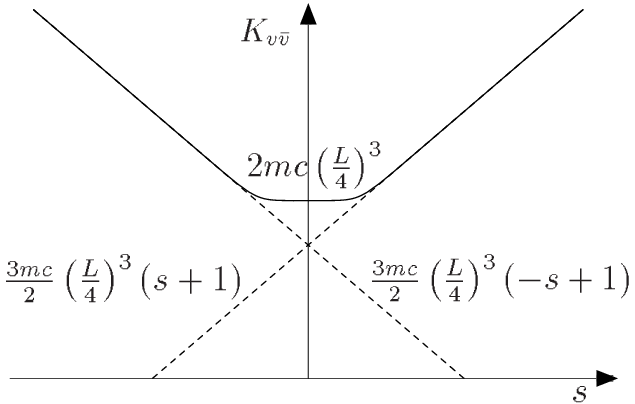}&
\includegraphics[width=7cm]{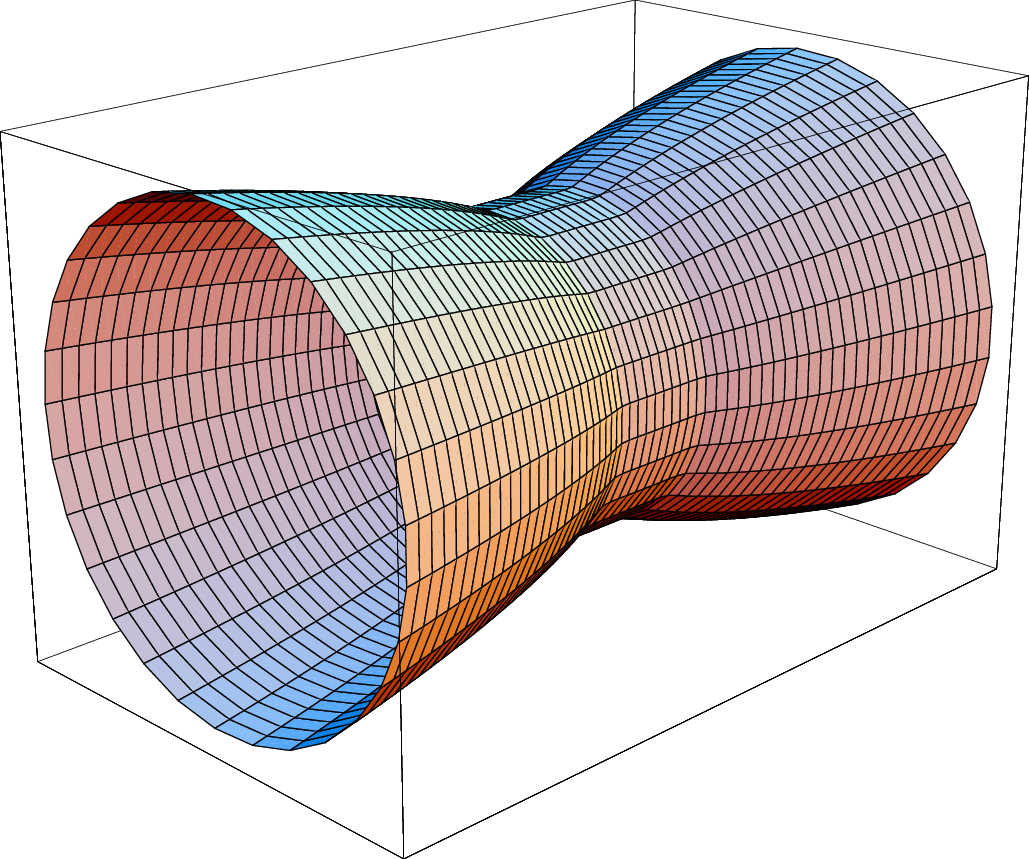}\\
(a) The metric of the moduli space & (b) The moduli space embedded into $\mathbf R^3$.
\end{tabular}
\caption{\small  
(a) The metric evaluated numerically (solid line) and the asymptotic metric (\ref{quad-wall}) computed from kinetic energies (dotted lines). 
(b) Embedding of the moduli space into the 3-dimensional Euclidean space. Here the metric is numerically evaluated in the limit $g \rightarrow \infty$. The moduli space is non-singular since the curvature is finite everywhere.}
\label{fig:embed-non}
\end{center}
\end{figure}
In the outer two regions of the parameter $s$, 
the metric has linear dependence on $s$ 
since the lengths of the internal walls 
depend linearly on $s$. 
We have observed the same feature 
in the case of the triangle loop in Eq.\,(\ref{eq:tri-asym}). 
In the middle region, the linear dependence on $s$ cancels out 
and the metric does not depend on $s$.
Fig.\,\ref{fig:embed-non}-(a) shows the 
the numerically evaluated metric and
Fig.\,\ref{fig:embed-non}-(b) shows the shape of the moduli space 
isometrically embedded into the 3-dimensional Euclidean space. 

\begin{figure}[htb]
\begin{center}
\includegraphics[width=7cm]{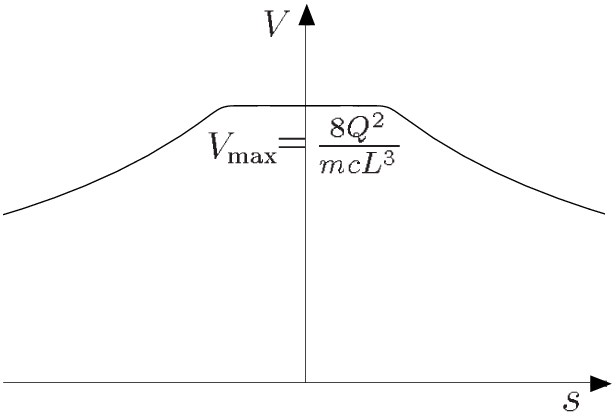}
\caption{\small The potential $V=\frac{Q^2}{4K_{v \bar v}}$.}
\label{fig:potential-non}
\end{center}
\end{figure}

If we consider the motion without flavor charge 
from the outer regions of the parameter $s$ 
to the direction of middle region, 
it goes through the middle regions 
and continue to go to the same direction. 
This motion corresponds to the motion of the loop changing 
the vacuum region inside the loop. 
If the configuration has non-zero flavor charge $Q \not = 0$, 
the potential term $V=\frac{Q^2}{4K_{v \bar v}}$ 
will be induced in the effective Lagrangian. 
The typical form of the potential is 
shown in Fig.\,\ref{fig:potential-non}. 
If the energy $E$ is greater than 
$V_{\rm max} \equiv \frac{8Q^2}{mcL^3}$, 
the change of the vacua inside the loop can occur 
as in the case of $Q = 0$. 
However, if $E<V_{\rm max}$, 
the quadrangle loop bounces back to be larger 
without changing the vacua inside the loop. 

\begin{figure}[htb]
\begin{center}
\begin{tabular}{ccc}
\includegraphics[width=4cm]{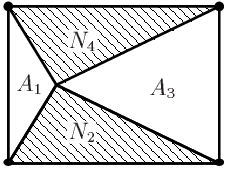}\quad&
\includegraphics[width=4cm]{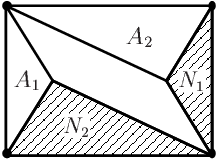}\quad&
\includegraphics[width=4cm]{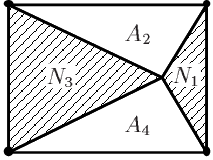}\\
$s<-1/3$&$-1/3<s<1/3$&$1/3<s$
\end{tabular}
\caption{\small The magnitude of the junction charge is proportional to
the area of triangle in the complex 
${\rm Tr}\langle \Sigma \rangle$ plane which is dual to the
junction point in the actual configuration. 
Abelian junctions $A_i$ and non-Abelian junctions $N_i$ 
$(i=1, \cdots, 4)$ are illustrated in Fig.\ref{fig:para}. 
Non-Abelian junctions are denoted by shaded regions. }
\vspace*{-.3cm}
\label{fig:junction}
\end{center}
\end{figure}

Let us next compute the kinetic energies of junctions. 
The magnitude of the junction charge 
is proportional to the area of 
the corresponding triangle in the grid diagram 
in the complex ${\rm Tr}\langle \Sigma \rangle$ plane. 
See equation Eq.\,(\ref{eq:abelian}) and Eq.\,(\ref{eq:non-abelian}). 
We show the areas of four junctions in Fig.\,\ref{fig:junction}.
The sign of the Abelian junction is minus 
while that of the non-Abelian junction is plus. 
Although the total junction charge 
is zero in all three regions of the parameter $s$, 
the positions of junctions have different dependence on $s$. 
This causes the different velocities of junctions 
for a given value of $\frac{ds}{dt}$, 
and the total kinetic energies of junctions can be non-zero. 
Since the Abelian junction transforms into 
the non-Abelian junction and vice versa at $s=\pm 1/3$, 
the kinetic energies of junctions 
are different in these three regions. 
These kinetic energies imply the additional contributions to 
the asymptotic metric as
\beq
\begin{array}{ccll}
ds_{\rm j}^2&=&\displaystyle-\frac{3}{8g^2}\left( \frac{L}{4} \right)^2
\Delta_{[134]}\,(ds^2+d \theta^2),&\hs{10} s \gg \displaystyle \frac{1}{3}, \\
ds_{\rm j}^2&=& \displaystyle 0, \phantom{-\frac{3}{8g^2}\left( \frac{L}{4} \right)^2
\Delta_{[134]}\,(ds^2+d \theta^2)}&\hs{10} \displaystyle s \approx 0, \\
ds_{\rm j}^2&=&\displaystyle\frac{3}{8g^2}\left( \frac{L}{4} \right)^2
\Delta_{[134]}\,(ds^2+d \theta^2),&\hs{10} s \ll - \displaystyle \frac{1}{3}.
\end{array}
\label{quad-wall-j}
\eeq

\section{Size Modulus Stabilization and $Q$ Webs of Walls}
\label{sc:stabilization}

So far, we have seen 
that the sizes of the loops tend to become larger 
after sufficiently long time interval. 
In this section, we show that the sizes of the loops 
stabilize if the third mass parameters $M_3$ are turned on 
in the Lagrangian (\ref{eq:lag}). 
One way to introduce $M_3$ consistently with supersymmetry 
is the Scherk-Schwarz 
dimensional reduction from 1+3 dimensions to 1+2 dimensions for
the Lagrangian (\ref{eq:lag}). 
Then the third (twisted) mass parameter $M_3$ is naturally 
introduced together with the third adjoint scalar 
$\Sigma_3$ whose origin is the gauge field of the reduced 
dimension. 

\subsection{Effective theory analysis}

Let us consider the triangle loop discussed 
in section \ref{section:triangle} for simplicity. 
If we turn on a small third mass parameter such that 
$M_3 = {\rm diag} \left(0,0,0,m_3 \right)$, 
a potential $V_m(r)$ is induced in the effective theory. 
We can show that this potential can be obtained 
from the 1+1 dimensional effective theory 
$\mathcal L = K_{w \bar w} 
\left(\p_\mu r \p^\mu r + \p_\mu \theta \p^\mu \theta \right)$ 
by requiring $r(t,x^3)=r(t),~ \theta(t,x^3) = \theta(t) + m_3 x^3$ 
and then reducing to 1-dimensional theory as 
\beq
L = K_{w \bar w}(r) \left[\left(\frac{dr}{dt}\right)^2 
+ \left(\frac{d\theta}{dt}\right)^2 \right]
- V_m(r), \hs{10} V_m(r) = (m_3)^2 K_{w \bar w}(r).
\eeq 
The asymptotic form of this potential 
obtained from the
asymptotic K\"ahler metric (\ref{eq:tri-asym}) takes the form
\beq
V_m(r) = (m_3)^2 \frac{c}{2\Delta_{[123]}} 
\frac{r}{\alpha_1 \alpha_2 \alpha_3}, 
\eeq
which is valid for $r \gg 1$. 
This is a confining 
potential so that the loop shrinks and eventually shrinks 
to a point, if we do not turn on the flavor $Q$-charges 
coming from the motion of the phase. 
This potential can be interpreted as a 
shift of energies of the walls. 
Because of the small mass parameter $m_3$, 
the tensions of walls shift as\footnote{
The Scherk-Schwarz dimensional reduction just introduces 
one more component of the energy density for the tension 
$T_{\rm w}$ of domain walls in Eq.(\ref{topo}). 
The tension of domain walls is formally still given by the 
same formula, proportional to the length of the mass vector: 
$T^{\left<A,B\right>} = c |\vec m_A -  \vec m_B|$, 
except that the mass vector $\vec m_A$ now becomes a 
three-vector in the three-dimensional grid diagram, 
after the Scherk-Schwarz dimensional reduction. 
}
\beq
\Delta T^{\langle A , 4 \rangle} ~=~ 
\sqrt{(T^{\langle A , 4 \rangle})^2 
+ (m_3 c)^2} - T^{\langle A , 4 \rangle} ~\approx~ 
\frac{(m_3c)^2}{2T^{\langle A , 4 \rangle}}.
\eeq
Therefore the total shift of energy can be evaluated 
by using Eq.\,(\ref{eq:tention-length}) as 
\beq
\sum_{A=1}^3 \Delta T^{\langle A , 4 \rangle} l^{\langle A , 4 \rangle} = (m_3)^2 \frac{c}{2\Delta_{[123]}} \frac{r}{\alpha_1 \alpha_2 \alpha_3}.
\eeq 

\begin{figure}[htb]
\begin{center}
\includegraphics[width=7cm]{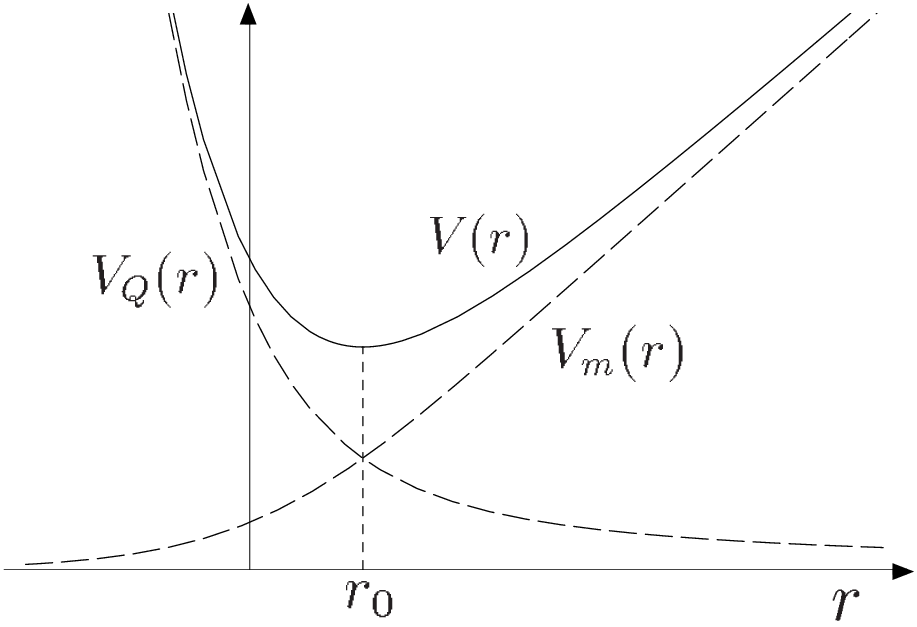}
\caption{\small The potential $V$ for non-zero third mass 
parameter $m_3$ and flavor charge $Q$. The potential $V$ 
is the sum of $V_m=(m_3)^2 K_{w \bar w}$ and 
$V_Q=\frac{Q^2}{4K_{w \bar w}}$.
The potential takes the minimum $|m_3Q|$ at $r=r_0$. 
}
\label{fig:V}
\end{center}
\end{figure}

When we turn on the flavor charge $Q$ around the domain walls
composing the loop given in Eq.~(\ref{eq:Q-charge}), 
the effective Lagrangian can be rewritten as 
\beq
&&\widetilde L = K_{w \bar w}(r) \, 
\left(\frac{dr}{dt}\right)^2 - V_Q(r) - V_m(r), \\
&&V_Q(r) + V_m(r) = 
\frac{Q^2}{4K_{w \bar w}} + (m_3)^2 K_{w \bar w} \ge |m_3 Q|.
\label{eq:pot_effctive}
\eeq
Let us remember that $K_{w\bar w}$ vanishes in the limit 
of $r\rightarrow -\infty$ 
and diverges in the limit of $r\rightarrow \infty$.
Therefore, $V_Q(r) \equiv \frac{Q^2}{4K_{w \bar w}}$ 
increases as $r \rightarrow - \infty$ 
while $V_m(r) \equiv (m_3)^2 K_{w \bar w}$ 
increases as $r \rightarrow \infty$ asymptotically, 
and there is the minimum of the potential saturating 
the inequality in the last equation,
as shown in Fig.\,\ref{fig:V}. 
At the minimum, the value of the K\"ahler potential is 
related to the given $Q$-charge
\beq
K_{w\bar w}= \frac{1}{2}\left|\frac{Q}{m_3}\right|.
\label{eq:minima_eff}
\eeq
Comparing this with 
the $Q$-charge for the unstable 
configuration in Eq.~(\ref{eq:Q-charge}), 
we observe that the stable configuration has 
$d\theta/dt = \pm m_3$. 
The relation (\ref{eq:minima_eff}) implies the size of the 
loop is stabilized at a certain value $r=r_0$. 
For sufficiently small 
value of $|m_3|$ 
and large value of $|Q|$, 
the stabilized size $r_0$ takes a large value, 
so that it can be evaluated from the asymptotic potential as 
\beq
r_0 = \frac{|Q|}{|m_3|c} \Delta_{[123]} \alpha_1 \alpha_2 \alpha_3. 
\label{eq:minimal_orbit}
\eeq

Thus the third mass parameter $M_3$ stabilizes the size 
moduli of loops by preventing the loops to expand forever. 
When the flavor charge $Q$ becomes non zero, the 
configuration is stabilized at a finite size of the loop, 
rather than at the boundary of the moduli space 
corresponding to the complete shrinkage (vanishing size) 
of the loop. 
This stabilization mechanism is the same 
as the one of the $Q$-lumps in nonlinear sigma models 
with a potential term \cite{Leese:1991hr}--\cite{Eto:2005sw}; 
the size of $Q$-lumps are stabilized 
by the $Q$-charge and the masses. 

In the subsequent section we will show the BPS nature 
of this stabilized loop configuration
from the viewpoint of original theory.
There we will see the origin of
the minimum of the potential 
(\ref{eq:pot_effctive}) which is always positive
except for the case $Q=0$ or $m_3=0$. 
The value of the potential $|Q m_3|$ at the minimum 
is shown to be equal to an increase in
the BPS mass from the BPS loop with $Q=m_3=0$.

\subsection{$Q$-domain wall web as 1/4 BPS soliton}

So far we have seen dynamics of $Q$ charged domain walls 
and their networks mainly from the 
viewpoint of the low energy effective theory. 
Let's go back to the original theory and reanalyse the 
stable network with non-zero $Q$ charges in more detail. 
It turns out that the configuration is a solution of 
another 1/4 BPS equations which are deformed 
from Eqs.\,(\ref{eq:BPS2}) 
and (\ref{eq:BPS1}). 
In order to see it, let us 
consider supersymmetric model 
in $d=2+1$ with 8 supercharges in which there is an 
additional adjoint scalar $\Sigma_3$ and third mass 
parameter $M_3$
as mentioned above. 

It is convenient to write the mass matrix as 
$M_3 \equiv \mathrm m_3^a \mathrm H_a = {\rm diag} \, 
(m_3^1,m_3^2,\cdots,m_3^{\NF})$, where we set 
${\rm Tr}\,M_3=0$ without loss of generality and 
$\mathrm H_a~(a=1,2,\cdots,\NF-1)$ are the generators 
of $U(1)^{\NF-1}$, 
that is, the elements of the Cartan subalgebra of $SU(\NF)$. 
The densities of conserved charges of the $U(1)^{\NF-1}$ 
symmetries are defined by
\beq
\rho_a \equiv i 
\left(H^1 \, \mathrm H_a \left(\D_0 H^1 \right)^\dagger 
- \D_0 H^1 \, \mathrm H_a H^{1\dagger}  \right).
\label{eq:Q-density}
\eeq
In addition, it is convenient to define an electric charge density as
\beq
\rho_e \equiv \p_\alpha {\rm Tr} \left( F_{0\alpha} \Sigma_3 \right).
\eeq
Then the energy density can be written as 
\beq
\mathcal E &=& {\rm Tr} \bigg[ \frac{1}{g^2} 
F_{\alpha0}^2 + \frac{1}{g^2} F_{12}^2 
+ \frac{1}{g^2} 
\left(\D_0 \Sigma_{\tilde\alpha}\right)^2 + \frac{1}{g^2} 
\left(\D_\alpha \Sigma_{\tilde\alpha}\right)^2 
- \frac{1}{2g^2}[\Sigma_{\tilde\alpha},
\Sigma_{\tilde\beta}]^2 \notag \\
&{}& + \phantom{\bigg[} |\D_0 H^1 |^2 
+ |\D_\alpha H^1|^2 + |H^1 M_{\tilde\alpha}
-\Sigma_{\tilde\alpha} H^1|^2 + \frac{g^2}{4} 
\left( H^1H^{1\dagger} - c 
\mathbf 1_{\NC} \right)^2 \bigg] \notag \\
&=& \!\!\phantom{\bigg[} {\rm Tr} 
\Bigg[ \frac{1}{g^2} \left( F_{12} 
- i [\Sigma_1,\Sigma_2] \right)^2 
+ \phantom{\bigg[} \frac{1}{g^2} 
\left( \D_1 \Sigma_1 + \D_2 \Sigma_2 - \frac{g^2}{2} 
\left( c \mathbf 1_{\NC} 
- H^1H^{1\dagger} \right) \right)^2 \notag \\
&{}& + \!\!\phantom{\bigg[} \frac{1}{g^2} 
\left( \D_1 \Sigma_2 - \D_2 \Sigma_1 \right)^2 
+ |\D_\alpha H^1 - \left(H^1 M_\alpha-\Sigma_\alpha H^1 \right) |^2 
\notag \\
&{}& + \frac{1}{g^2} \left( F_{\alpha0} 
+ \D_\alpha \Sigma_3 \right)^2 
+ \frac{1}{g^2} \left( \D_0 \Sigma_\alpha + i [\Sigma_3, \Sigma_\alpha] \right)^2 
+ \frac{1}{g^2} \left( \D_0 \Sigma_3 \right)^2 \notag \\
&{}& + \phantom{\bigg[} |\D_0 H^1 
- i \left(H^1 M_3-\Sigma_3 H^1 \right)|^2 \bigg] 
+ \mathcal Y + \mathcal Z_1 + \mathcal Z_2 
+ \mathrm m_3^a \rho_a + \frac{2}{g^2} \rho_e + \p_\alpha J_\alpha \notag \\
&\geq& \mathcal Y + \mathcal Z_1 + \mathcal Z_2 
+ \mathrm m_3^a \rho_a + \frac{2}{g^2} \rho_e + \p_\alpha J_\alpha,
\label{eq:Q_bound}
\eeq
where 
$\alpha$ stands for indices $1,2$ while $\tilde\alpha$ 
for $1,2,3$, and 
we have used the Gauss's law 
\beq
\D_\alpha F_{\alpha0} - i[\Sigma_{\tilde\alpha}, \D_0 \Sigma_{\tilde\alpha}]
- i \frac{g^2}{2} \left(H^1 \D_0 H^{1\dagger} - \D_0 H^1 H^{1\dagger} \right) = 0.
\label{eq:Gauss}
\eeq 
The BPS equations are obtained by requiring the BPS bound 
to be saturated. 
Apart from the equations to determine time-dependence, we 
find the same five 1st order equations as 
Eqs.\,(\ref{eq:BPS2}) and (\ref{eq:BPS1}) 
for the fields $\{W_\alpha,\Sigma_\alpha, H^1\}$ 
\beq
&F_{12} - i [\Sigma_1,\Sigma_2] = 0, 
\hs{5} \D_1 \Sigma_2 - \D_2 \Sigma_1=0, 
\hs{5} \D_1 \Sigma_1 + \D_2 \Sigma_2 
= \frac{g^2}{2} \left( c \mathbf 1_{\NC} - H^1H^{1\dagger} \right),
& \notag \\
&\D_1 H^1 - H^1 M_1 + \Sigma_1 H^1 = 0, 
\hs{5} \D_2 H^1 - H^1 M_2 + \Sigma_2 H^1 = 0.&
\eeq
Then the solution, except for time dependence, can be 
written by using the solution of the master equation 
Eq.\,(\ref{eq:master}) $\Omega \equiv SS^\dagger$ as 
Eq.\,(\ref{eq:solBPS1}) and Eq.\,(\ref{eq:solBPS2}). 
Time dependence of all the fields including new 
variables $\{W_0,\Sigma_3\}$ are determined by 
additional equations \cite{Eto:2005sw} 
\beq
&F_{\alpha0} + \D_\alpha \Sigma_3 = 0, 
\hs{5} \D_0 \Sigma_\alpha + i [\Sigma_3, \Sigma_\alpha] = 0, 
\hs{5} 
\D_0 \Sigma_3 = 0, \notag \\
&\D_0 H^1 - i H^1 M_3 + i \Sigma_3 H^1 = 0.&
\eeq
If we choose a gauge such that $W_0 = - \Sigma_3$, 
these can be solved by replacing $H_0$ as 
$H_0 \rightarrow H_0 \, e^{i M_3 t}$ and requiring the 
other fields to be independent of time
\beq
i W_1 + \Sigma_1 = S^{-1} \p_1 S, 
\hs{5} i W_2 + \Sigma_2 = S^{-1} \p_2 S, 
\hs{5} H^1 = S^{-1} H_0 \, e^{M_1 x^1 + M_2 x^2 + i M_3 t}.
\label{eq:solQ-web1}
\eeq
The spatial profile of the gauge field $W_0$ 
(and adjoint scalar $\Sigma_3$) are finally determined 
from the Gauss's law constraint Eq.\,(\ref{eq:Gauss}). 
The solution takes the form of 
\beq
W_0 = - \Sigma_3 = - \mathrm m_3^a 
\displaystyle \left( \frac{\partial}{\partial \sigma^a} 
S^\dagger S^{\dagger-1} + S^{-1} 
\frac{\partial}{\partial \bar{\sigma}^a} S \right).
\label{eq:solQ-web2}
\eeq
Here, $\sigma^a\,(a=1,2,\cdots,\NF-1)$ are the complex 
moduli parameters whose imaginary parts correspond to the 
Nambu-Goldstone modes of $U(1)^{\NF-1}$ symmetries, 
that is, $\sigma^a$ appear in the expression of $H_0$ 
as $H_0(\sigma^a) = H_0(\sigma^a=0)\,e^{\sigma^a \mathrm H_a}$. 
It is instructive to note a similarity 
between Eqs.~(\ref{eq:solQ-web2}) 
and (\ref{eq:sol_Gauss}). Recall that the latter is 
the solution of the Gauss's law constraint for the 
configurations when we promote the moduli parameters 
of the stationary 1/4 BPS background to fields 
(functions of the world-volume coordinates including 
time $t=x^{\mu=0}$). 
Let us suppose that we make an Ansatz for the field 
$\sigma^a$ to depend only linearly on 
time as $\sigma^a(t)=\sigma^a + i\mathrm m_3^at$, 
namely the moduli matrix changes as 
$H_0 \, e^{\sigma^a \mathrm H_a} \to H_0 \, 
e^{\mathrm H_a (\sigma^a + i \mathrm m_3^at)}$. 
This leads to 
$\partial_0 \sigma^a(t) = i \mathrm m_3^a$ 
and then the solution (\ref{eq:sol_Gauss})
corresponds to the solution (\ref{eq:solQ-web2}). 

For the solution Eqs.\,(\ref{eq:solQ-web1}) and 
(\ref{eq:solQ-web2}), the $Q$-charges are determined by 
integrating the densities $Q_a = \int d^2 x \, \rho_a$. 
Interestingly, the $Q$-charge which is a variable defined 
in the original theory can be directly related to the 
K\"ahler metric of the low energy effective theory 
(\ref{eq:general}) 
\beq
Q_a = 2 \mathrm m_3^b K_{a\bar b} = 2 \mathrm m_3^b K_{b \bar a}, 
\hs{10} 
K_{a\bar b}\equiv\frac{\partial}{\partial \bar \sigma^b} 
\frac{\partial}{\partial \sigma^a} K = K_{b \bar a}.
\label{eq:Q_K}
\eeq
Note that the K\"ahler potential $K$ is 
independent of the imaginary parts of $\sigma^a$.
Since the right-hand side depends on the parameters 
contained in $H_0$, some of these parameters 
are fixed for given values of $Q_a$. 
Therefore, those parameters are
no longer moduli parameters and the configuration 
is stabilized. 
Especially, in the case of the triangle loop discussed in 
the previous subsection, 
we can show that the size parameter is fixed at the same 
value obtained as the minimum of the potential 
Eq.\,(\ref{eq:minima_eff}) in the effective theory 
by taking, for example, 
$\sigma^a = \left(0, 0, \sqrt{\frac{3}{2}}w\right)$ and 
$\mathrm m_3^a = \left(0, 0, \sqrt{\frac{3}{2}} m_3\right)$ 
with $\mathrm H_{a=3} = \frac{1}{2\sqrt6}{\rm diag}(-1,-1,-1,3)$ 
and the notation 
$Q = \sqrt{\frac{3}{2}} Q_{a=3}$. 
Furthermore, the 
minimum value $|m_3 Q|$ of the effective potential 
in Eq.~(\ref{eq:pot_effctive}) precisely corresponds to the 
increment of energy bound $\mathrm 
m_3^a Q_a$ in the 
last line of Eq.~(\ref{eq:Q_bound}). 
Note that since $F_{0\alpha} \rightarrow 0$ at spatial infinity in this case, the electric charge does not contribute to the total energy
\footnote{
The Q-wall can be viewed as a capacitor with electric charge distributions on the two sides of the wall \cite{Bolognesi:2007nj}. Since these charges have opposite signs, the total electric charge vanishes.
}.
Eq.~(\ref{eq:Q_K}) tells us that the 1/4 BPS configuration 
requires both non-zero $Q$-charges and the third masses 
$\mathfrak m_3^a$ (or both of them to vanish simultaneously). 
If one of them is absent, the balance between them is lost 
and the configuration no longer is BPS. 
This is also consistent with what we found from the 
effective theory viewpoint 
in the previous sections.

\section{Conclusion and Discussion}

In this paper, we investigated dynamics of 1/4 BPS 
domain wall networks (or webs)
in Abelian or non-Abelian gauge theories coupled with 
complex masses for 
Higgs fields in the 
fundamental representation. 
In the previous paper~\cite{Eto:2006bb} we 
have obtained the effective action on the world-volume 
of the domain wall networks. 
In this paper we 
applied it to study the dynamics of the networks. Namely, 
We described the dynamics of the slowly moving networks 
as geodesics on their moduli space, namely 
with the moduli approximation. 
Only moduli parameters related to internal loops composed of 
several domain walls 
in the networks can be treated 
as massless fields in the effective Lagrangian. 
Other moduli are associated with the shift of 
external domain walls, which requires an infinite amount 
of energy and results in the change of boundary conditions.

As concrete examples, we dealt with three different types 
of loops in Abelian or non-Abelian gauge theories. 
The first example is in Sec.~\ref{section:triangle} where 
the simplest configuration of the single triangle loop 
appears in the Abelian gauge theory with 4 massive Higgs fields. 
The metric of the moduli space for the single
loop has a geometry between a cone and a cigar~\cite{Eto:2006bb}.
We found geodesics corresponding 
to any motion of a shrinking 
loop pass the tip (zero size of the 
loop). 
This means that the loop bounces back with $\pi$ 
rotation of the internal phase 
and eventually expands 
to an infinite loop.
The second example is the dynamics of two loops in the 
Abelian gauge theory with 
6 massive Higgs fields in Sec.~\ref{section:double}. 
There exist seven 
types of configurations shown 
in Fig.~\ref{fig:configurations}.
We numerically showed that after sufficiently long time 
both of the two loops expand forever 
irrespective 
of the 
initial condition. 
As the two loops get larger, the system approaches 
to a system of two independent single loops. 
Our last example is the network including 
both Abelian and non-Abelian junctions which appears 
in the $U(2)$ gauge theory with 4 massive Higgs 
fields~\cite{Eto:2005fm}, see Sec.~\ref{non-abe}. 
After fixing all the non-normalizable moduli, 
there remains only one complex moduli parameter $s$ 
as the normalizable modulus which controls the areas of two 
vacuum regions. 
We found the metric of the moduli space whose geometry 
 in Fig.~\ref{fig:embed-non} looks like a sandglass 
made by gluing the tips of the 
two metrics of a single triangle loop in 
Fig.~\ref{fig:embed-tri}. 
The geodesic of $s$ is a one-way traffic from any initial 
value to an expanding loop with either $s=+\infty$ or 
$s=-\infty$ (one or the other branches of the sandglass), 
depending on its initial velocity. 
Namely only 
one of the loop out of two loops remains after 
sufficient time.

We also considered the dynamics of the web loop accompanied 
by a phase rotation in 
the internal direction, a $U(1)$ isometry which originates 
from a linear combination of broken $U(1)$ flavor symmetries.
Conserved charges associated with the rotation are 
$Q$-charges of the domain walls composing the loops.
These $Q$-charges give 
a runaway potential in the effective theory and 
exert a repulsive force between walls in the loop. 
Then the loops with $Q$-charges are generally unstable 
(non-BPS) and tend to expand 
forever. 
Thanks to the repulsive force, 
the geodesics bounce back before reaching the 
completely shrunk loops. 
The minimum sizes of the loops are determined by the 
given $Q$-charges.
For the loops including both the Abelian and the 
non-Abelian junctions with the sandglass geometry 
in Fig.~\ref{fig:embed-non}, 
the corresponding geodesic motion becomes 
bounce-back type or one-way traffic type 
depending on the total energy and the 
given $Q$-charge.

By introducing the third masses for the Higgs fields, 
the effective theory of the loops acquires an attractive 
potential in contrast to 
the $Q$-charges. 
In the presence of the third masses, 
the loops tend to shrink. 
Coexistence of the $Q$-charges and the third 
masses stabilize the size of the loops. 
Then the size of the loop is fixed 
at some value where the attractive and the repulsive 
forces are balanced like the known stabilization mechanism of 
size moduli for the lumps due to the suitable 
potential accompanied by the 
$Q$-charge~\cite{Leese:1991hr}--\cite{Eto:2005sw}. 
We also studied such configurations in the original theory, 
rather than in the effective theory on the world volume of 
the web of loops. 
Then we derived new 1/4 BPS equations which includes 
time derivatives and found 
a new BPS bound which is the sum of 
the topological charges of domain walls and their junctions 
and the 
additional masses coming from the $Q$-charges. 
General solutions of the 1/4 BPS equations with the 
Gauss's law are found. 
All the results found in the original theory are compatible 
with those found in the effective theory.

Here we make several comments on 
possible extensions of the present work. 

Global cosmic strings appear when 
global $U(1)$ symmetry is spontaneously broken.  
It is well-known that there exists 
a repulsive force between two strings. 
Strings interact with the Nambu-Goldstone boson  
associated with the spontaneously broken $U(1)$ symmetry,  
and the repulsive force was explained 
in terms of 
the Nambu-Goldstone bosons 
propagating in the bulk \cite{VS}. 
In the same way, it will be possible to explain 
the repulsive force induced inside a domain wall loop 
in terms of the 
Nambu-Goldsonte bosons. 
A feature 
different from the case of cosmic strings is that 
the Nambu-Goldstone modes in this case of a domain wall loop 
are normalizable and therefore appear in 
the low energy effective action of the loop 
as shown in this paper.

An 
extension to a supertube \cite{Mateos:2001qs} 
is interesting and may have 
some impact on string theory. 
With a Noether charge density,  
domain walls with arbitrary shape \cite{Kim:2006ee} 
were constructed as a field theory realization 
of a supertube. 
Our work should be extendible to a BPS supertube junction. 
Such a solution may suggest 
a junction of membrane tube in M-theory.

It is interesting to explore applications of our results 
to cosmology.  
Our theory is supersymmetric and 
all stationary configurations discussed in this paper are BPS 
and are stable. 
In the early Universe supersymmetry is 
expected to be unbroken. 
Therefore our results can be applied  
when gauge and global symmetry are broken 
above the supersymmetry breaking scale. 
Vacuum regions inside domain wall loops are 
considered to be bubbles. 
Our results imply all bubbles grow 
for a late time in theory with complex masses 
if the size of 
the Universe is infinite. 
In this respect, it is worth generalizing 
our work to domain wall webs 
in a finite size space. 
In this case, zero modes of 
external legs of walls become normalizable and 
are promoted to fields in the effective theory. 
Dynamics of webs is not restricted to loops 
and will become 
richer.
If we do not restrict ourselves to supersymmetric Universe, 
we can allow 
triplet masses of Higgs fields  
even in four space-time dimensions. 
In this case, the bubble (loop) sizes are stabilized. 
Growing bubbles with (without) 
a $Q$-charge will be stabilized (shrink)
after supersymmetry is broken and the triplet masses 
are induced.

\section*{Acknowledgments}
This work is supported in part by Grant-in-Aid for 
Scientific Research from the Ministry of Education, 
Culture, Sports, Science and Technology, Japan No.17540237
and No.18204024 (N.~S.). 
The work of T.F.~is 
supported by the Research Fellowships of the Japan Society for
the Promotion of Science for Young Scientists.
The work of M.E.~and K.O.~is also supported by the Research Fellowships of the Japan Society for
the Promotion of Science for Research Abroad.
T.~N. gratefully acknowledges 
support from a 21st Century COE Program at 
Tokyo Tech ``Nanometer-Scale Quantum Physics" by the 
Ministry of Education, Culture, Sports, Science 
and Technology, 
and 
support from the Iwanami Fujukai Foundation.


\appendix 

\section{Triangle loop with degenerate masses}\label{appendix:degenerate}

Let us consider a triangle loop for $\NC=1, \NF=5$ case 
in which the masses for fourth and fifth flavor components 
are degenerate $\vec m_4 = \vec m_5$. 
There exist four Killing vector fields $\xi_0,\,\xi_a~(a=1,2,3)$ 
on the moduli space, which are given by 
\beq
\xi_0 \equiv i  \phi^i \p_i + \mbox{(c.c.)}, 
\hs{5} \xi_a \equiv i \left( \sigma_a \right)_{ij} 
\phi^j \p_i + \mbox{(c.c.)}.
\eeq
These Killing vectors originate from the $U(2)$ flavor 
symmetry which rotates the fourth and fifth flavor components of $H$. 
Not all of them are independent. 
Instead they  
are related as
\beq
\xi_0 = \frac{1}{|\boldsymbol \phi |^2} 
\left( \bar \phi \sigma_a \phi \right) \xi_a.
\eeq
The tangent space of the moduli space  
can be orthogonally decomposed into the direction of 
size of the loop $t_r \equiv \phi^i \p_i + \mbox{(c.c.)}$, 
phase of the loop $t_\theta \equiv \xi_0 $ and two 
directions of the vacuum moduli inside the loop 
$t_I \equiv c_I^a \xi_a~(I=1,2)$. Here the coefficients 
$c_I^a$ are defined by $c_I^a 
\left( \bar \phi \sigma^a \phi \right) = 0,~c_I^a c_J^a 
= \delta_{IJ}$. 
The norms of these vector fields are given by
\beq
&{}&\|t_r\|^2 = \|t_\theta\|^2 
= 2|\boldsymbol \phi |^2 \Big( K'(|\boldsymbol \phi|^2) 
+ |\boldsymbol \phi|^2 K''(|\boldsymbol \phi|^2) \Big) 
= \frac{1}{2} \frac{\p^2}{\p r^2} K, \label{eq:norm1}\\
&{}&\|t_1\|^2 = \|t_2\|^2 
= 2|\boldsymbol \phi |^2 K'(|\boldsymbol \phi|^2)
= \frac{\p}{\p r} K,\label{eq:norm2}
\eeq
Here $r \equiv \log |\boldsymbol \phi|$ can be 
interpreted as the size of the loop.
For large $r$, 
$K \rightarrow \frac{c}{3 \Delta_{[123]}} 
\frac{1}{\alpha_1 \alpha_2 \alpha_3}r^3$ 
and the norms Eq.\,(\ref{eq:norm1}), (\ref{eq:norm2}) become
\beq
\|t_r\|^2 = \|t_\theta\|^2 ~\rightarrow~ \displaystyle 
\frac{c}{\Delta_{[123]}} \frac{r}{\alpha_1 \alpha_2 \alpha_3}, \hs{10}
\|t_1\|^2 = \|t_2\|^2 ~\rightarrow~ 
\displaystyle \frac{c}{\Delta_{[123]}} 
\frac{r^2}{\alpha_1 \alpha_2 \alpha_3}.\label{eq:norm3}
\eeq
These asymptotic form of the norms and their 
dependence 
on the size of the loop $r$ show the fact that 
the metric densities for size and phase moduli have 
a one-dimensional support on the edges of the loop, 
while those for the vacuum moduli have a two-dimensional 
support extended fully inside the loop
(See Fig.\,\ref{fig:intra}). 
Since the moduli space has isometrics generated by the Killing vectors $\xi_0$ and $\xi_a$, there are conserved Noether charges defined by
\beq
Q &=& \langle \dot \phi,\, \xi_0 \rangle ~=~ i K_{i \bar j} \left( \dot{\bar \phi}^j \phi^i - \dot \phi^i \bar \phi^j \right), \\
q_a &=& \langle \dot \phi,\, \xi_a \rangle ~=~ i K_{i \bar j} \left( \dot{\bar \phi}^j {\left( \sigma_a \right)^i}_k \phi^k - \dot \phi^i \bar \phi^k {\left( \sigma_a \right)_k}^j \right),
\eeq
where $\dot \phi \equiv \dot \phi^i \p_i + \mbox{(c.c.)}$ and $\langle \,,\, \rangle$ denotes the inner product with respect to the metric of the moduli space $K_{i \bar j}$.
These conserved charges are related as
\beq
Q = \frac{1}{|\boldsymbol \phi |^2} \left( \bar \phi \sigma_a \phi \right) q_a.
\eeq
Since the tangent vectors $t_r,t_\theta,t_1,t_2$ are orthogonal, the time derivative of the moduli parameters $\dot \phi$ can be wirtten as
\beq
\dot \phi = \frac{\langle \dot \phi ,\, t_r \rangle}{\|t_r\|^2} t_r + \frac{\langle \dot \phi,\, t_\theta \rangle}{\|t_r\|^2} t_\theta + \frac{\langle \dot \phi,\, t_1 \rangle}{\|t_1\|^2} t_1 + \frac{\langle \dot \phi,\, t_2 \rangle}{\|t_2\|^2} t_2. 
\eeq
Then the energy in the effecitve theory can be written as
\beq
E ~=~ \frac{1}{2} \langle \dot \phi ,\, \dot \phi \rangle &=& \frac{1}{2} \left( \frac{\langle \dot \phi ,\, t_r \rangle^2}{\|t_r\|^2} + \frac{\langle \dot \phi ,\, t_\theta \rangle^2}{\|t_\theta\|^2} + \frac{\langle \dot \phi ,\, t_1 \rangle^2}{\|t_1\|^2} + \frac{\langle \dot \phi ,\, t_2 \rangle^2}{\|t_2\|^2}\right) \notag \\
&=& \frac{1}{4} \p_r^2 K \, \dot r^2 + \frac{Q^2}{\p_r^2 K} + \frac{1}{2 \p_r K} \left(c^a_1 c^b_1 + c^a_2 c^b_2 \right) q_a q_b \notag \\
&=& \frac{1}{4} \p_r^2 K \, \dot r^2 + \left(\frac{1}{\p_r^2 K}-\frac{1}{2\p_r K} \right) Q^2 + \frac{q_a q_a}{2 \p_r K}.
\eeq
Here we have used $\langle \dot \phi,\, t_r \rangle = \dot r \, \|t_r \|^2$ and $c^a_1 c^b_1 + c^a_2 c^b_2 + (\bar \phi \sigma^a \phi)(\bar \phi \sigma^b \phi)/|\boldsymbol \phi|^4 = \delta^{ab}$. 

\end{document}